\documentclass[10pt,a4paper]{article}
\pdfoutput=1

\usepackage[utf8]{inputenc}
\usepackage[T1]{fontenc}

\usepackage{amsmath}
\usepackage{amssymb}
\usepackage{bbm}
\usepackage{graphicx}
\usepackage{authblk}
\usepackage{siunitx}

\title{Geometry of bounded critical phenomena}
\author[1,2,3,*]{Giacomo Gori}
\author[3,4,5]{Andrea Trombettoni}
\affil[1]{Institut für Theoretische Physik, Universität Heidelberg, D-69120 Heidelberg, Germany}
\affil[2]{Dipartimento di Fisica e Astronomia ``Galileo Galilei'' Università di Padova, Via Marzolo 8, I-35131 Padova, Italy}
\affil[3]{CNR-IOM DEMOCRITOS Simulation Center, Via Bonomea 265, I-34136 Trieste, Italy}
\affil[4]{SISSA and INFN, Sezione di Trieste, Via Bonomea 265, I-34136 Trieste, Italy}
\affil[5]{Dipartimento di Fisica, Università di Trieste, Strada  Costiera 11,  I-34151  Trieste,  Italy}
\affil[*]{Corresponding author: gori@sissa.it}

\date{\today}

\begin{document}
\maketitle

\begin{abstract}
The quest for a satisfactory
understanding of systems
at criticality in dimensions
$d>2$ is a major field of
research.
We devise here a geometric description of 
bounded systems at criticality
in any dimension $d$.
This is achieved by altering 
the flat metric with a space dependent
scale factor $\gamma(x)$, $x$ belonging 
to 
a 
bounded domain $\Omega$.
$\gamma(x)$ is
chosen in order to have a 
scalar curvature to be
constant and matching the one of the
hyperbolic space, the proper 
notion of curvature being -- as
called in the mathematics literature --
the fractional Q-curvature.
The equation for $\gamma(x)$
is found to be the 
Fractional Yamabe Equation (to be solved in $\Omega$)
that, in absence of anomalous
dimension, 
reduces to the
usual Yamabe Equation in the same domain. 
From the 
scale factor $\gamma(x)$ we obtain novel 
predictions for the scaling
form of 
one-point order parameter correlation functions.
A (necessary) virtue of the
proposed approach is that
it encodes and allows to
naturally retrieve the purely
geometric content of
two-dimensional boundary conformal
field theory. 
From the critical magnetization
profile in presence of
boundaries one
can extract the scaling dimension of the order parameter, $\Delta_\phi$.
For the 3D Ising model
we find $\Delta_\phi=0.518142(8)$
which favorably compares 
(at the fifth decimal place)
with the state-of-the-art estimate.
A nontrivial prediction
is the structure of two-point spin-spin
correlators
at criticality. They should 
depend on the fractional 
Q-hyperbolic distance
calculated from the
metric, in 
turn depending only on the
shape of the bounded domain
and on $\Delta_\phi$.
Numerical simulations
of the 3D Ising model
on a slab geometry
are found to be in 
agreement with such predictions.
\end{abstract}

\maketitle

\section{Introduction}
The understanding of critical phenomena and 
critical states 
is a central theme of modern physics.
In the course of the study of criticality
powerful tools have been developed 
such as the Renormalization Group 
able to enlighten universal physical properties from
condensed matter systems to high
energy physics.

A basic concept that has emerged
as a defining property of the 
critical state is that of being \emph{scale
invariant}. We refrain from
working out these concepts
in full generality~\cite{cardy} while
concentrating on the case
of a bounded system defined
in a domain $\Omega$ topologically
equivalent to a ball.
It is well known that,
at criticality, for a wide 
range of systems~\cite{cardy,mussardo} 
one can adopt a continuum
description where the physical
observables depend on the spatial
coordinate, in our case $x\in\Omega\subset\mathbb{R}^d$.
Consider a (space dependent) 
observable 
depending on the local degrees of
freedom $s(x)$ located inside the domain,
think of the order parameter 
$\phi(x)=s(x)$
as a guiding example.
In the critical state the system
will have a given distribution
of the microscopic degrees of
freedom allowing 
to obtain
averages that will be denoted by $\langle\ldots\rangle$.
If the system admits a well
defined thermodynamic limit
the averages are expected to
 converge 
to some well behaved function $\langle \phi(x) \rangle$.

The consideration of bounded systems 
has both practical and theoretical
advantages. On the practical side
when dealing with real systems
(be them experimental or numerical
realizations) we are in general 
treating finite systems.
On the theoretical side boundary
theories are often more
constraining leading to 
more refined results, moreover corrections
to the infinite system
system behavior contain information
on some fundamental properties
of the theory such as the 
central charge~\cite{difrancesco,cardyc}. These relations
allow us to access these fundamental
quantities even in finite systems.

Scaling hypothesis for the 
operators implies that if we take
a system $\lambda \Omega$
of a size $\lambda$ 
times larger than $\Omega$
then the correlation function of operator $\phi$
will be given by:
\begin{equation}
\langle \phi(\lambda x)\rangle_{{}_{\lambda\Omega}}
 = \lambda^{-\Delta_\phi} \langle \phi (x)\rangle_{{}_{\Omega}}
\end{equation}
where $\Delta_\phi$ is the scaling 
dimension of the field $\phi$~\cite{cardy,fisher}.
Similar Ans\"atze can be put forward for 
observables depending on more points.

\section{Uniformisation}
We now put forward our main working hypothesis (\verb|Uniformisation|):\\

\emph{A system at criticality in a bounded domain will try to
modify its (flat euclidean) metric 
in order to be ``as uniform
as possible''.}\\ 

Making the above statement precise and derive from it quantitative,
testable nontrivial predictions that will be checked constitutes the aim of the present
work. The flat metric
will be denoted by $\delta=\delta_{ij}$
where the indices $i$ and $j$
run from $1$ to $d$;
the symbol $g$ will instead be
reserved for a generic
metric $g=g_{ij}$.
The allowed change in
the metric will be of the type
$\delta \rightarrow \delta/\gamma(x)^2$
where $\gamma(x)$ is a space
dependent scale factor.
A reason for allowing changes 
of this type is that on short
scales, such that the effect of boundaries
is negligible, the system should locally behave
as a bulk system which is isotropic. 
Such a change in the metric is known 
in the mathematical literature as
a \emph{conformal change} and 
metrics related by these transformations
are said to belong to the same conformal class.

Since we are trying to
set up an intrinsic geometry, the uniformisation
should entail curvatures. As we are aiming at fixing 
one space dependent function $\gamma(x)$
we will have to put constraints
on 
a quantity. A first reasonable guess
(which will be later modified) is the scalar curvature, $R_g$, where
we explicitly noted the dependence on
the metric $g$. For convenience we
remind how (a normalised version of) $R_g$ is defined in terms of the
metric $g_{ij}$:
\begin{align}
\Gamma_{jk}^i&=\frac12 g^{il} \left( \partial_{k} g_{lj} + \partial_{j} g_{lk} - \partial_{l} g_{jk} \right)\\
 \mathrm{Ric}_{ij} &=
\partial_{l}{\Gamma^l_{ji}} - \partial_{j}\Gamma^l_{li}
+ \Gamma^l_{l\lambda} \Gamma^\lambda_{ji}
- \Gamma^l_{j\lambda}\Gamma^\lambda_{li}\\
R_g&=\frac{1}{d(d-1)}\mathrm{Ric}_{ij} g^{ji}\label{normricci}
\end{align}
(as usual, summation over repeated indices
is assumed).

Thus, according to uniformisation, we would end up
with the equation:
\begin{equation}
 R_{\delta/\gamma(x)^2} = \kappa.
\end{equation}
Actually for two-dimensional systems
this guess, implementing our uniformisation
hypothesis, appears to be well motivated
since scalar
curvature alone is a quantity fully specifying
geometric properties~\cite{docarmo}. Now it comes to the choice 
of the right constant value $\kappa$ to set for $R$.
We have the following possibilities:
\begin{itemize}
 \item $\kappa>0$ is constant positive curvature, think of the sphere $\mathbb{S}^d$ as a (very special) example, which does not appear to be suited to describe
 a system with boundary since it has no borders.
 \item $\kappa=0$ is a flat space (actually the one we started with) which is also not suited to pursue Uniformisation since the points living near the boundary cannot be
 treated on the same footing as the other points in the bulk.
 \item $\kappa<0$ is constant negative curvature, think of the hyperbolic space $\mathbb{H}^d$ as a (again very special) example, indeed appears as a reasonable since it is endowed with an infinitely distant boundary.
\end{itemize}
Since one can always rescale $\kappa$, from now on
we choose $\kappa=-1$\footnote{From time to time
the constant $\kappa$ is restored, keep
in mind however that it should always be set to $-1$.}.

Writing down the equation requires the knowledge 
of the transformation laws of scalar curvature 
under conformal changes of the metric.
We have that 
$R_{g/\gamma(x)^2} = R_{g} - |\nabla \gamma(x)|^2 +\frac{2}{d} \gamma(x) \Delta \gamma(x)$.
The problem we have just stated is 
known as the Yamabe problem~\cite{yamabe}. 
Since the $g$ we are starting from
is flat ($R_g=0$) and we are requiring $R_{g/\gamma(x)^2}=\kappa=-1$ ,
we are aiming at the solution of:
\begin{equation}\label{YE}
 1 - |\nabla \gamma(x)|^2 +\frac{2}{d} \gamma(x) \Delta \gamma(x) = 0,
\end{equation}
i.e. the so-called Yamabe Equation. In~\eqref{YE}
gradient and Laplacian are calculated with the flat metric
and the function $\gamma$ should be zero on $\partial\Omega$, the boundary 
of $\Omega$.
The Yamabe problem, an old acquaintance to geometers, 
is the subject of extensive  mathematical research in
the literature at the interface of analysis and 
geometry~\cite{schoen,lee,maryamabe}.
The Yamabe Equation for $d=2$ has been studied 
in connection with the Liouville field theory~\cite{zamzam} 
and for $d>2$ it just occasionally surfaced in the
physics literature~\cite{kiessling, kholodenko}.

Another more suggestive form of the above equation is written
as the nonlinear eigenvalue problem for the positive definite operator $(-\Delta)$:
for $d\neq 2$
\begin{equation}\label{NLEYE}
 (-\Delta) \gamma(x)^{-\frac{d-2}{2}} = -\frac{d(d-2)}{4}\gamma(x)^{-\frac{d+2}{2}}.
\end{equation}
For $d=2$ a limit has to be performed yielding the Liouville equation
\begin{equation}
  (-\Delta) \log \gamma(x) = -\kappa \gamma(x)^{-2}.
\end{equation}
In $d=2$ the solution of the above problem inside a (connected and simply connected)
domain $\Omega$ amounts exactly to the construction of a model of hyperbolic
space $\mathbb{H}^2$. 

We quote two simple solutions of (\ref{NLEYE})
valid in any $d$: putting $x=(x_1, x_2,$ $\ldots, x_d)$, 
we have {\it i)} for the upper half hyper-space, 
$x_d>0$, it is $\gamma(x)=x_d$; {\it ii)} for a ball of radius $r$ 
we have $\gamma(x)=\frac{r^2-x^2}{2 r}$ where $x^2=\sum_{i=1}^d x_i^2$ . These two are examples where
by conformally altering the metric we can construct
$\mathbb{H}^d$, i.e. a space where not only $R$ but 
all the sectional curvatures are constant.
The two-dimensional case stands on its own
because as already mentioned $R$ is enough
to specify the geometric properties
of the space. The spaces constructed
inside say a disk and a square will be isometric
and the coordinate change between them 
will be the conformal mapping between
the interiors of the square and the disk 
(so in this case a Schwarz-Christoffel mapping)
that due to Riemann mapping theorem exists,  
provided the domains are regular enough.

But let's pursue the geometric reasoning. Having 
obtained a uniformising metric, we wish to
construct from it predictions for observables.
The obtained metric indeed constitutes, locally,
a gauge for measuring lengths. One-point correlators
should be function of it. Let us now inspect
how the solution of the Yamabe Equation changes under
a rescaling of the domain. We have that
\begin{equation}
 \gamma_{{}_{\lambda \Omega}}(\lambda x) = \lambda \gamma_{{}_{\Omega}}(x).
\end{equation}
Given this transformation law, based on our
uniformisation hypothesis, we are led to conjecture that:
\begin{equation}\label{conj1pt}
 \langle \phi(x) \rangle = const. \times \gamma(x)^{-\Delta_\phi},
\end{equation}
where $\gamma$ is the solution of Yamabe Equation~\eqref{YE}.
We anticipate that in \eqref{conj1pt} we are not
fully taking into account the effect of
anomalous dimension on the metric; this will be fixed
in the next two Sections.

We now turn to two-point correlators.
For the two-point correlator $ \langle \phi(x) \phi(y) \rangle$, a 
prefactor $\gamma(x)^{-\Delta_\phi} \gamma(y)^{-\Delta_\phi}$ 
restoring the correct physical dimensions is expected,
while what is missing, because of our 
hypothesis 
of a purely geometric description, 
should be only a function $F$ of
the distance $\mathfrak{D}_{\delta/\gamma^2}(x,y)$ between
points $x$ and $y$, calculated with the metric 
$\delta/\gamma^2$. This yields:
\begin{equation}\label{conj2pt}
 \langle \phi(x) \phi(y) \rangle = \gamma(x)^{-\Delta_\phi} \gamma(y)^{-\Delta_\phi}
 F(\mathfrak{D}_{\delta/\gamma^2}(x,y)).
\end{equation}
The above considerations can be extended for higher order correlators.
For example three-point functions should contain three dimensional
prefactors and an arbitrary function of three mutual distances and so on.

What we have just outlined is indeed true for 
two-dimensional systems and, with $d>2$, for  
systems defined in the upper half hyperspace 
or inside a hypersphere as derived by
using the group of conformal symmetries~\cite{cardyboundary}.
In $d=2$, for one-point functions
it coincides with known results in bounded critical systems,  
see e.g.~\cite{cardyparplat, burkhardt, straley, BCFT},
while, always in $d=2$, for higher point correlators
it matches 
the transformation
law for correlators in boundary conformal
field theories (see e.g. 5.24 in~\cite{difrancesco}) 
under conformal mappings. From our geometric
viewpoint this is traced back to the fact that all 
complete, connected, simply-connected
spaces of constant negative scalar curvature
are isometric in two dimensions.

In order to see explicitly in $d>2$ that our results 
agrees with symmetry-based derivations~\cite{cardyboundary}, 
recall that for the $d$-dimensional upper half hyperspace the
hyperbolic distance is $\mathfrak{D}_{\delta/x_d^2} (x,y)
=\mathrm{arccosh} \left[1+\frac{|x-y|_{d-1}^2+(x_d-y_d)^2}{2 x_d y_d}\right]$
where $|x-y|_{d-1}^2=\sum_{i=1}^{d-1} (x_i-y_i)^2$.
Our conjecture for two-points then
exactly 
reproduces what can be found in formula 
(3.9) of~\cite{cardyboundary}.
We stress that this is special to
the upper half hyperspace and inside
a hypersphere domains that do not 
acquire a dependence on anomalous
dimension. Our results refer instead 
to any domain in any dimension. For domains 
different from the upper half space and the
ball the anomalous dimension plays a role in $d>2$, as we discuss in 
the next Section.

We remark another rewarding property of the 
structure of solutions of Yamabe problem:
close to the boundary of $\Omega$, as it can 
be gleaned from \eqref{YE}, we have that
$\gamma(x)$ is proportional to the euclidean
distance to the boundary $\partial\Omega$.
This implies a locality property: near the boundary
the system effectively looks
like a hyperbolic space forgetting
about the detailed shape of the domain.
This feature is also retained
by solutions of the 
fractional Yamabe problem
that will shortly be introduced.

The explicit analytical solution for the Yamabe Equation 
in a slab domain relevant for the
interpretation of numerical
experiments is presented in Appendix~\ref{YamabeAppendix}.

\section{Anomalous dimensions inclusion}
The content of this section
is not meant to give a microscopic
derivation of our main 
results, expressed by~\eqref{conj1ptFYE}-\eqref{conj2ptFYE},
but to provide qualitative
arguments to support and motivate
the conjectures that will be
stated in Section~\ref{confcovform}
and verified in Section~\ref{compnumexp}.

Let us reconsider the Yamabe Equation
in a different light. 
Take a general theory which is
at most
quadratic in the fields. Its action
is given by:
\begin{equation}
 \mathcal{S}[\phi(x)] = \int \mathrm{d}^d x \frac{1}{2} \phi(x) 
 (A \phi)(x) - \int \mathrm{d}^d x b(x) \phi(x),
\end{equation}
where $A$ is a general linear operator which should
be positive in order to ensure control over fluctuations
of the field. A good example is obviously minus 
the Laplacian, $A=(-\Delta)$. $b(x)$ is an external magnetic field. 
With this action we can perform 
averages as follows:
\begin{equation}
 \langle \phi(x') \phi(x'') \phi(x''') \ldots \rangle_Q = \frac{1}{\mathcal{Z}} \int \mathcal{D}[\phi(x)] (\phi(x') \phi(x'') \phi(x''') \ldots) \exp(-\mathcal{S}[\phi(x)])
\end{equation}
with the normalization $\mathcal{Z} = \int \mathcal{D}[\phi(x)] \exp(- \mathcal{S}[\phi(x)])$.
We can calculate the average of the order parameter obtaining
\begin{equation}
 \langle \phi(x)\rangle_Q = A^{-1} b(x)
\end{equation}
where the inverse of $A$ has appeared. Inverting the above relation we get
\begin{equation}\label{eqQ}
A \langle \phi(x)\rangle_Q = b(x).
\end{equation}
Let us now take $A=(-\Delta)$ and pursue some geometric
considerations based on scaling. Assume there is a metric, $\gamma_Q (x)$, 
describing the system at criticality. The field $\phi$ has scaling
dimensions $\frac{d-2}{2}$, also known as free field or canonical
dimensions,  
while $b(x)$ will have scaling dimensions 
$\frac{d+2}{2}$ for the action to be dimensionless~\cite{cardy,mussardo}.
With these assumptions Eq.~\eqref{eqQ} reads:
\begin{equation}
(-\Delta) \gamma_Q(x)^{-\frac{d-2}{2}} = const.\times \gamma_Q(x)^{-\frac{d+2}{2}}
\end{equation}
and it has reproduced
the Yamabe Equation.
We stress again that this is not meant to be a rigorous derivation,
but a way to obtain equations with correct dimensional properties.\footnote{A 
rigorous derivation would start from an interacting theory
based on the scale invariant action $\mathcal{S}_I$
constructed with the order parameter field $\phi$:
$$
\mathcal{S}_I[\phi(x)] = \int \mathrm{d}^d x \frac{1}{2} \phi(x) 
 (-\Delta) \phi(x) + g_c \int \mathrm{d}^d x \left(\phi(x)^2\right)^\frac{d}{d-2}
$$
where $g_c$ is a coupling constant.}

Let us now exploit and insert some common knowledge we have from
the theory of critical phenomena. A generic
observable will have some scaling dimension
differing
from the free field
one. 
The presence of so-called \emph{anomalous}
dimensions is at the heart of
the existence of a nontrivial 
theory of critical phenomena. 
As an example take the magnetization
in the 3D Ising model which has a scaling
dimension of $\Delta^{CB}_\phi = 0.5181489(10)$~\cite{boot3}. 
We quote the best result to date obtained via the Conformal Bootstrap 
technique~\cite{boot3}, other high precision determinations 
of $\Delta_\phi$ are to be found in~\cite{boot1,boot2,Hasenbusch0,ferrenberg} 
and are reported in Table~\ref{table1}
presented in Section~\ref{compnumexp}. 
This result differs by a small, but definitely nonzero 
amount from the expected canonical 
dimension $\frac{d-2}{2}=\frac{1}{2}$. 

How can we correct
the above equations to account for the
anomalous scaling? A first guess 
would be to simply substitute the order parameter
with $\gamma(x)^{-\Delta_\phi}$ and 
the conjugate field with $\gamma(x)^{-d+\Delta_\phi}$.
However the Laplacian would have the 
wrong scaling dimensions. A natural way out
is to consider a power of it:
\begin{equation}\label{FYEQ}
(-\Delta)^{\frac{d}{2}-\Delta_\phi} \left[\gamma_{(\Delta_\phi)}(x)\right]^{-\Delta_\phi} 
= const.\times \left[\gamma_{(\Delta_\phi)}(x)\right]^{-d+\Delta_\phi},
\end{equation}
where the subscript $_{(\Delta_\phi)}$ on $\gamma$ signals the dependence 
on the anomalous dimension $\Delta_\phi$. Eq.~\eqref{FYEQ} is the so-called
Fractional Yamabe Equation.\footnote{{
In this case a candidate interacting theory from which a derivation (possibly formal) 
of the fractional Yamabe equation could be attempted should have a scale 
invariant action entailing the fractional laplacian.
}} 
It
emerged in the context of the fractional Yamabe
problem, i.e. finding metrics making
generalizations of the scalar curvature, 
the so-called fractional $Q$-curvatures, 
constant. The first definition of
this problem with rigorous results for the case of compact manifolds
appeared in~\cite{mar_quing}.

We remind that in the full space $\mathbb{R}^d$ (with
no boundaries) several definitions of the
fractional Laplacian are known in literature~\cite{pozrikidis,maryamabe} 
and they {\it do} coincide~\cite{kwasnicki}.
However Equation \eqref{FYEQ} is to be solved
in the bounded domain $\Omega$.
When adapted to bounded domains, the
different definitions of the 
fractional Laplacian in general
{\it do not} any longer coincide
and this is both a problem for 
applications and a challenge 
for mathematical research
which is currently subject 
of intense work~\cite{lischke}. 

For the purposes of the present
paper we anyway need to define and solve Equation \eqref{FYEQ}
in bounded domains. This has to be done in
order to compare the results
with lattice Monte Carlo simulations to
validate our Uniformisation hypothesis
and the conjectures, that will be stated in the following 
Equations~\eqref{conj1ptFYE}-\eqref{conj2ptFYE},
for one-point and two-point correlations.

The route we follow to overcome these difficulties
is to introduce a conformally
covariant version of the fractional Laplacian
in bounded domains, as detailed in
the next section.

For later convenience we introduce $s$,
the order of the fractional Laplacian: $$s=\frac{d}{2}-\Delta_\phi.$$

\section{Conformally covariant formulation}\label{confcovform}
While the derived equations~\eqref{NLEYE}
and its generalisation~\eqref{FYEQ} 
make sense, their appearance is not
so satisfying since they depend on
operators defined in the reference
(flat) space. It would be very
appealing to have operators 
transforming in a consistent way
under conformal changes of the
metric. Actually it is much more than
an aesthetic consideration,
since to have good conformal transformation properties
is a quite strong requirement
for the construction of a
fractional Laplacian.

On a manifold $M$ of dimension $d$
(this definition is usual 
for compact manifolds, while we shall
need it in the non-compact case), 
we define an operator $A_g$ to
be conformally covariant
if under a conformal change
in the metric $g\rightarrow g' = g/w^2$
($w(x)$ being an arbitrary positive gauge function)
the relation 
\begin{equation}
 A_{g'}(w^{p}\varphi) = w^{q} A_{g}(\varphi) 
\end{equation}
holds, where $\varphi$ is a function in $C^{\infty}(M)$
and $p$ and $q$ are constants.
Notably the operator
\begin{equation}
 \mathcal{L}^{(1)}_g = (-\Delta_g) + \frac{d(d-2)}{4}R_g,
\end{equation}
called the conformal Laplacian,
falls under this classification
with $p=\frac{d-2}{2}$ and $q=\frac{d+2}{2}$, 
being $-\Delta_g$ the Laplace-Beltrami
operator for the metric $g$ and $R_g$
its scalar curvature as defined in~\eqref{normricci}.
With this operator
the Yamabe Equation transforms into
\begin{equation}
\mathcal{L}^{(1)}_{\delta/{w(x)^2}} \left(\frac{\gamma(x)}{w(x)}\right)^{-\frac{d-2}{2}} = -\frac{d(d-2)}{4} \left(\frac{\gamma(x)}{w(x)}\right)^{-\frac{d+2}{2}}
\end{equation}
and takes an especially simple
form if we write it in terms of the metric $\delta/\gamma(x)^2$: 
\begin{equation}
\mathcal{L}^{(1)}_{\delta/\gamma(x)^2} (\mathbbm{1}) = -\frac{d(d-2)}{4} (\mathbbm{1}).
\end{equation}
This means that the metric uniformising 
scalar curvature is the one which acting
with its associated conformal Laplacian 
on a constant field (denoted by $\mathbbm{1}$) brings it to (a
multiple of) the constant field,
making in some sense  its uniformising
properties more explicit.
Of course the numerical difficulties
of solving the equation are still there.

Other operators (not general enough for this
work) are the conformally covariant 
integer powers of the Laplacian:
the Paneitz operator~\cite{branson} and the GJMS
operators~\cite{GJMS}. 
Most important are instead scattering
operators first defined in~\cite{graham}
for compact manifolds whose 
definition has been reconciled
with more conventional definitions
of the fractional Laplacian in~\cite{changmar}.

In terms of the properly defined fractional
Laplacian $\mathcal{L}^{(s)}_{g}$ 
of order $s$, Fractional Yamabe Equation reads:
\begin{equation}
\mathcal{L}^{(\frac{d}{2}-\Delta_\phi)}_{\delta/w^2} \left(\frac{\gamma_{(\Delta_\phi)}(x)}{w(x)}\right)^{-\Delta_\phi} = 
\frac{\Upsilon(\Delta_\phi)}{\Upsilon(d-\Delta_\phi)} \left(\frac{\gamma_{(\Delta_\phi)}(x)}{w(x)}\right)^{-d+\Delta_\phi}. 
\end{equation}
Again choosing $w(x)^2=\gamma_{(\Delta_\phi)}(x)^2$ one has
\begin{equation}\label{FYEcovariant}
\mathcal{L}^{(\frac{d}{2}-\Delta_\phi)}_{\delta/\gamma_{(\Delta_\phi)}^2} (\mathbbm{1}) = 
\frac{\Upsilon(\Delta_\phi)}{\Upsilon(d-\Delta_\phi)} (\mathbbm{1}), 
\end{equation}
where the dependence on $\Delta_\phi$
of $\gamma$ has been explicitly noted. 
With this notation the solution
of Yamabe Equation is $\gamma_{(\frac{d-2}{2})}(x)$.
The constant has been fixed 
in terms of the function 
${\Upsilon(x) \equiv \Gamma(1-x)\,\sin(\pi x/2)}$
such that the hyperbolic space $\mathbb{H}^d$ is a solution, proof
of this statement and calculation of the constant
are to be found in Appendix~\ref{scatteringAppendix}.
The two conjectures stated in the previous Section, 
Equations~\eqref{conj1pt}-\eqref{conj2pt},
stay the same, {\it but} with the anomalous 
dimension dependent scale factor $\gamma_{(\Delta_\phi)}(x)$.
Thus we can put forward the following\\
\vspace*{0.125cm}\\
{\bf Conjecture for one-point correlators:}
\begin{equation}\label{conj1ptFYE}
 \langle \phi(x) \rangle = const. \times \left[\gamma_{(\Delta_\phi)}(x)\right]^{-\Delta_\phi}, 
\end{equation}\\
\vspace*{0.125cm}\\
and\\
\vspace*{0.125cm}\\
{\bf Conjecture for two-point correlators:}
\begin{equation}\label{conj2ptFYE}
 \langle \phi(x) \phi(y) \rangle = 
 \left[\gamma_{(\Delta_\phi)}(x)\right]^{-\Delta_\phi} 
 \left[\gamma_{(\Delta_\phi)}(y)\right]^{-\Delta_\phi}
 F(\mathfrak{D}_{\delta/{\gamma_{(\Delta_\phi)}}^2}(x,y)).
\end{equation}\\
\vspace*{0.125cm}\\
Remember that $\mathfrak{D}_g(x,y)$
is the distance calculated with the metric $g$
and $\gamma_{(\Delta_\phi)}(x)$ is the solution of
Equation~\eqref{FYEcovariant}.

Equations~\eqref{conj1ptFYE} and \eqref{conj2ptFYE} are the main results of 
this paper, and we emphasize they are intended to be valid 
in any bounded domain 
at criticality in any dimension. We expect 
similar formulas to hold for higher-point correlators.

Predictions from~\eqref{conj1ptFYE}-\eqref{conj2ptFYE}
will be checked against numerical simulations 
for the Ising model in a non-trivial domain, the slab, in the next Section.
Notice that if one, unlike our approach, assumes at criticality 
just an
effective theory having only fractional derivatives
in the kinetic term (and no interaction terms),
one would face in a bounded domain the 
following two inherent problems: first
one would be forced to choose which
definition of the fractional Laplacian
has to be used, and moreover the 
structure of the different correlators
would obey unavoidably Wick theorem,
leaving no freedom for higher-point 
correlators.

Equipped with the operator $\mathcal{L}^{(s)}$, we can write down
an expression for the fractional Q-curvature, $R_g^{(s)}$, for a generic 
metric $g$:
\begin{equation}
\label{fracQR}
R_g^{(s)}=
-\frac{\Upsilon\left(\frac{d}{2}+s\right)}{\Upsilon\left(\frac{d}{2}-s\right)}
\mathcal{L}^{(s)}_{g} \left(\mathbbm{1}\right)
\end{equation}where the coefficient in front of the rhs is chosen in order to 
have the fractional Q-curvature to be minus one for $\mathbb{H}^d$.
In order to not interrupt the flow of the presentation, we 
give in the Appendices 
\ref{scatteringAppendix}-
\ref{spectralFYEAppendix} the details needed for the formal construction of the fractional
Laplacian $\mathcal{L}^{(s)}_{g}$ in a bounded domain and the numerical solution of  
the Fractional Yamabe Equation~\eqref{FYEcovariant} in the slab geometry. The construction relies
on considering the $d$-dimensional domain $\Omega$ as the boundary
of a suitably defined $d+1$ dimensional space~\cite{MarGori}. 
The obtained findings will be compared with 
lattice Monte Carlo simulations in the same geometry. 

\section{Comparison with numerical experiments}\label{compnumexp}

In order to test our predictions we will consider the fruit fly
of statistical mechanics: the Ising model~\cite{mussardo}. 
We perform Monte Carlo simulations 
on a slab geometry, where the value of the spins is fixed 
to the value $1$ on the two planes delimiting 
the slab. The model is simulated at the critical temperature.
The value of the magnetization in this geometry only 
depends on the distance from the planes,
and we record as well the two-point correlation
functions.
The profile of the magnetization near boundary at criticality 
for the Ising model and many other statistical
mechanics models has been thoroughly
investigated in the literature of boundary
critical phenomena~\cite{diehl}.
In particular the boundary conditions we are using
correspond to the so-called extraordinary phase transitions~\cite{diehl}.
In the thermodynamic limit the magnetization
has to be rescaled by multiplying it by 
$L^{\Delta_\phi}$ yielding -- at criticality -- a
universal scaling function~\cite{cardyboundary}. 
A crucial step to reduce 
finite size effects is 
to introduce the \emph{extrapolation length} $a$~\cite{Hasenbusch1}, 
accounting for the expected power-law 
divergence not occurring 
exactly at the boundary in the lattice
system. 
Collapse of numerical 
data can be used this way to obtain
estimates of $\Delta_\phi$~\cite{Hasenbusch1}, 
but the resulting value
for it is not especially precise.
Much better Monte Carlo
estimates for $\Delta_\phi$
are rather obtained
by analyzing data with cross correlations 
between various thermodynamic quantities~\cite{ferrenberg} 
or determining by finite size scaling the value of parameters 
where leading corrections
to scaling vanish~\cite{Hasenbusch0},
however reaching to date a 
significantly smaller
precision than 
the Conformal Bootstrap
estimates~\cite{boot3}.
A summary of the best 
results for $\Delta_\phi$
is in Table~\ref{table1}
together with our new
estimate, obtained as follows.

Our determination is based on the knowledge
of the solution $\gamma_{(\Delta_\phi)}$ of 
the Fractional Yamabe Equation for the slab geometry.
In accordance with our 
conjecture \eqref{conj1ptFYE} for one-point
operators we have
$\langle \phi(x) \rangle \propto \left[\gamma_{(\Delta_\phi)}(x)\right]^{-\Delta_\phi}$.
Denoting by $i$ the lattice coordinate
in the transverse direction 
of the slab, $i=0,\ldots L$, we compute
from Monte Carlo simulation the magnetization
$m_i=\langle s_i \rangle=\langle \phi(x) \rangle$ where $s_i$
is the discrete Ising variable and $x=\frac{2i}{L}-1$ so that 
$x\in[-1,1]$.

We then fit the magnetization data using:
\begin{equation}\label{onepointfit}
m_i = 
\alpha L^{-\Delta_\phi} \left[\gamma_{(\Delta_\phi)}\left(\frac{x}{1+a/L}\right)\right]^{-\Delta_\phi},
\end{equation}
where the parameters $a$, $\alpha$, and $\Delta_\phi$ are left
free. Since we want to determine $\Delta_\phi$ 
in an unbiased fashion (not relying on previous estimates) 
the function $\gamma_{(\Delta_\phi)}(x)$
has to be determined for a range a values, that is chosen to be $[0.5,0.54]$.
In Appendix~\ref{spectralFYEAppendix}, Figure~\ref{IYEandFYE}
a contour plot of this function
is reported. We see to our surprise that the dependence on $\Delta_\phi$
around the free field value $\Delta_\phi=0.5$
is pretty weak.

\begin{figure}
 \begin{center}
 \includegraphics[width=.85\textwidth]{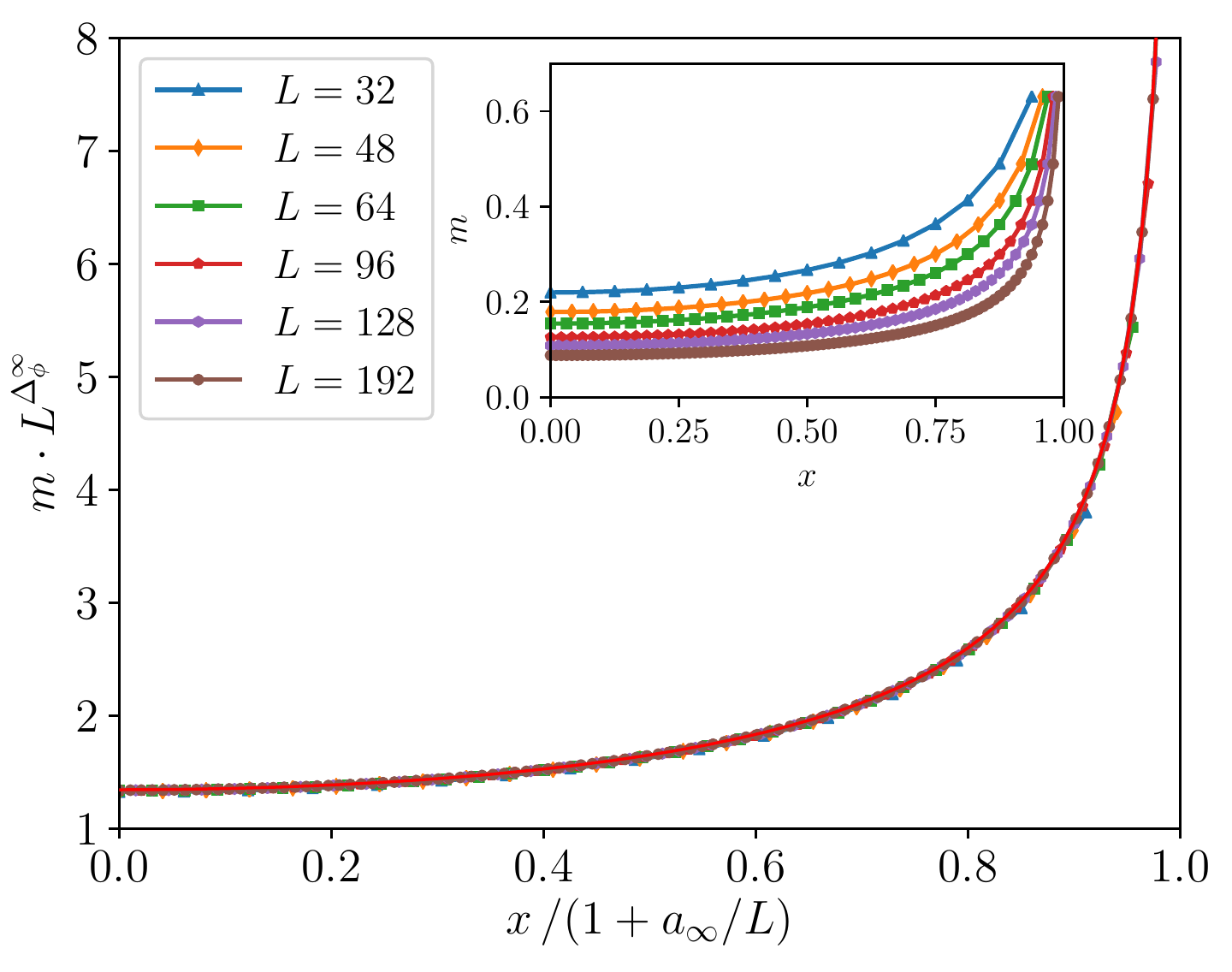}
 \end{center}
 \caption{Collapse plot of the magnetization data for
the different sizes considered. The parameters used, 
$a_\infty$, $\alpha_\infty$, $\Delta_\phi^{\infty}$,  
 are the ones extrapolated for $L=\infty$. The red line is the 
 universal scaling function $\alpha_\infty\left[\gamma_{(\Delta_\phi^\infty)}(x)\right]^{-\Delta_\phi^\infty}$ for the magnetization. The inset displays the raw data obtained from the simulations. In all cases 
 the errors are smaller than the size of the points.}\label{fig1}
\end{figure}

To obtain accurate results
it is important to minimize fine
size corrections to scaling. 
Because of universality, 
we are free to choose a model
within the same universality
class. Such a
model has already been devised~\cite{Hasenbusch0}
and used to obtain the most
refined Monte Carlo numerical results. 
It is the improved Blume-Capel model 
(at criticality), whose Hamiltonian 
is reported in Appendix~\ref{MCAppendix}
together with details of the
simulations. In order to assess
the validity of our predictions
the magnetization has been measured
and compared with our conjecture. The results for the magnetization 
are reported in Figure \ref{fig1}, where we also show the collapse 
of the data.

\begin{figure}
 \begin{center}
 \includegraphics[width=.85\textwidth]{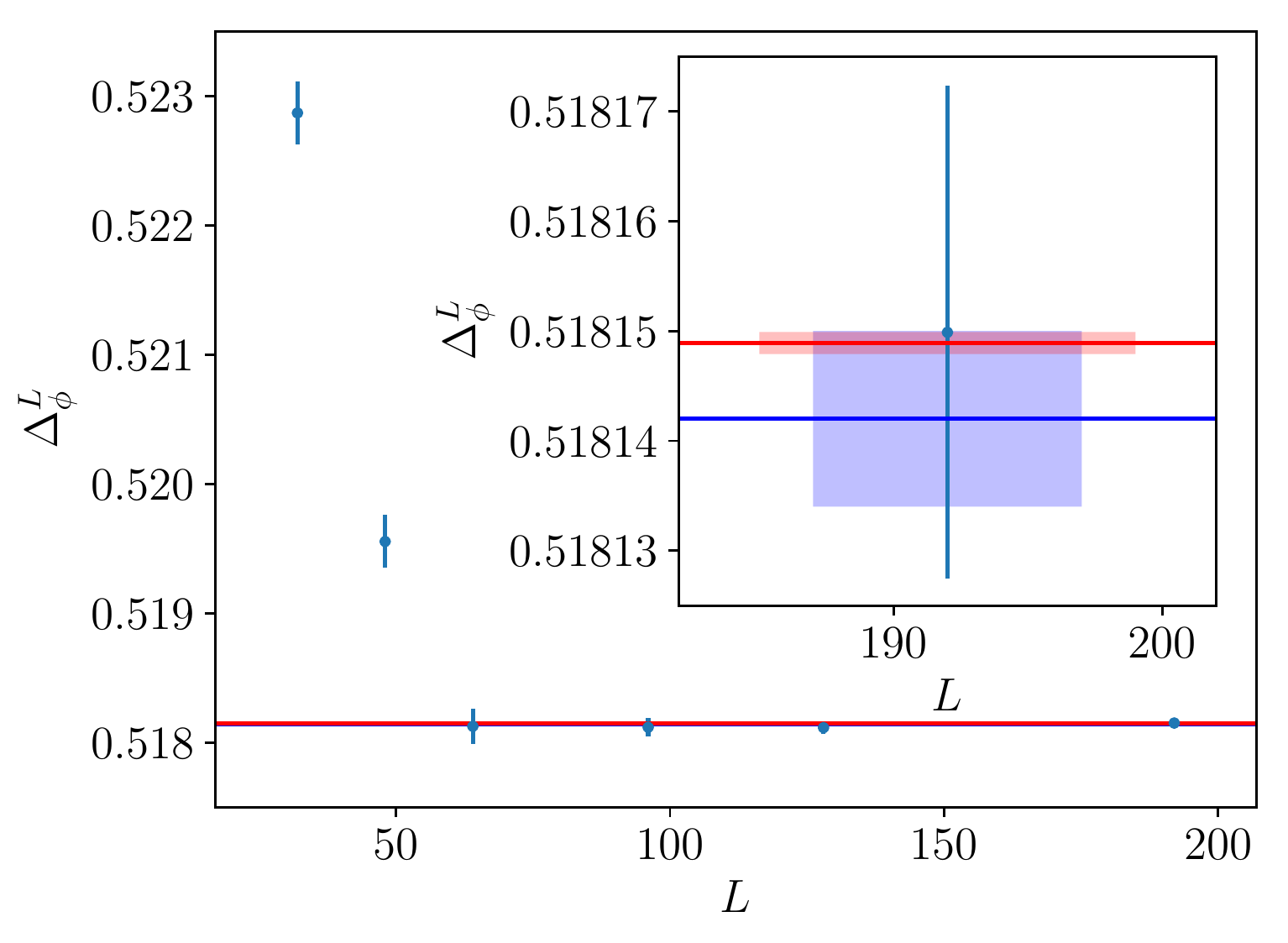}
 \end{center}
 \caption{Estimated values of $\Delta_\phi^L$ for the simulated sizes. 
The horizontal blue line represents our best estimate $\Delta_\phi^\infty$, 
while the red line is $\Delta_\phi^{CB}$~\cite{boot3}. 
The inset is a zoom for the largest 
size. In the main figure the error $\Delta_\phi^{CB}$ is reported 
as a shade. In the inset as well errors are represented as shades, 
however the shades are broken in order to better appreciate the 
overlap between the estimates.}\label{fig2}
\end{figure}

Details of the fitting 
procedure are in Appendix~\ref{dataAppendix}. The results  
for $\Delta_\phi$ are plotted in Figure \ref{fig2}. 
The obtained estimates are close to the best Conformal Bootstap result
available so far
$\Delta_\phi^{CB}$~\cite{boot3}, and the $L=192$ value, $\Delta_\phi=0.518150(22)$, 
is extremely close to it.
Since the data are all compatible 
with each other for $L\geq64$, 
we are allowed to perform 
a weighted average of them
yielding the value $\Delta_\phi=0.518142(8)$.
This value is compatible with $\Delta_\phi^{CB}$
and it has an order of magnitude 
larger error. In turn it is more precise, 
 by an order of magnitude 
than the best 
MC estimates reported
in the literature~\cite{Hasenbusch0,ferrenberg}.
All these results are summarised
in Table~\ref{table1} with our
estimate denoted by ``Critical Geometry''
for brevity. 

In Table~\ref{table2} we report the values obtained for the different sizes 
 in two ways. In the central column we use the metric 
$\gamma_{(\Delta_\phi)}$ obtained by solving the Fractional Yamabe Equation with a $\Delta_\phi$ which is 
left free and extracted from the fit of numerical results. In the right column 
we report the values obtained using the non-fractional Yamabe profile raised to a
power $\Delta_\phi$ left free. On one side one can observe that the latter values obtained 
from the Yamabe Equation are not accurate as much as the 
ones reported in the central column when compared with the best Conformal Bootstrap value 
$\Delta_\phi^{CB}$. On the other side the Yamabe Equation value is anyway rather 
good. The reason for this (somehow unexpected) result is the 
already mentioned weak dependence 
of the solution of the Fractional Yamabe Equation on the anomalous dimension, 
signaling the rigidity of the hyperbolic 
spaces whatever constant curvature is imposed. The anomalous 
dimension contribution to the metric factor is however 
important to get highly accurate estimates for $\Delta_{\phi}$. 
Anyway, we consider the fact that the 
non-fractional Yamabe Equation produces good results as a 
confirmation of the reliability of our geometrical approach. 
Therefore we expect 
that for more complicated domain shapes, where the 
solution of the Fractional Yamabe 
Equation may be very difficult to find, one could use the integer Yamabe 
Equation as a first good approximation. 

\begin{table}
\centering
    \begin{tabular}{l|l|l} 
      \textbf{Reference} & \textbf{Method} & \textbf{$\qquad\Delta_\phi$}\\
      \hline
      Hasenbusch\,(2010) \cite{Hasenbusch0}& MC & 0.518135(50)\\
      Ferrenberg et al.\,(2018) \cite{ferrenberg}& MC & 0.51801(35)\\
      Sheer El-Showk et al.\,(2014) \cite{boot2} & Conformal Bootstrap & 0.518154(15)\\
      Kos et al.\,(2016) \cite{boot3} & Conformal Bootstrap & 0.5181489(10)\\
      This paper & Critical Geometry & 0.518142(8)
    \end{tabular}
\caption{Best results for the 3D Ising scaling exponent 
of the order parameter $\Delta_\phi$. 
The best results to our knowledge to date for $\Delta_\phi$ are contained 
in lines 1 and 4 
pertaining to MC and Conformal Bootstrap method respectively. The last 
line is the value obtained using the approach described in this paper and denoted as
``Critical Geometry''.}
\label{table1}
\end{table}

\begin{table}
\centering
    \begin{tabular}{l|l|l} 
      \textbf{Linear size $L$} & \textbf{$\Delta_\phi$ 
      FYE profile fit} & \textbf{$\Delta_\phi$ 
      YE profile fit}\\
      \hline
      32 & 0.52287(24)   & 0.52570(17) \\
      48 & 0.51955(21)   & 0.52200(15) \\
      64 & 0.51812(13)   & 0.52038(7)\\
      96 & 0.51812(7)    & 0.51983(3)\\
      128 & 0.51811(5)   & 0.51931(3)\\
      192 & 0.518150(22) & 0.518923(15)\\
    \end{tabular}
\caption{Size dependent fitting values for $\Delta_\phi$ using (central column) 
the Fractional Yamabe Equation (FYE) profile
and (right column) the ordinary Yamabe Equation (YE)
profile for different 
values of the linear size $L$ (left column).}
\label{table2}
\end{table}

\begin{figure}
 \begin{center}
 \includegraphics[width=.77\textwidth]{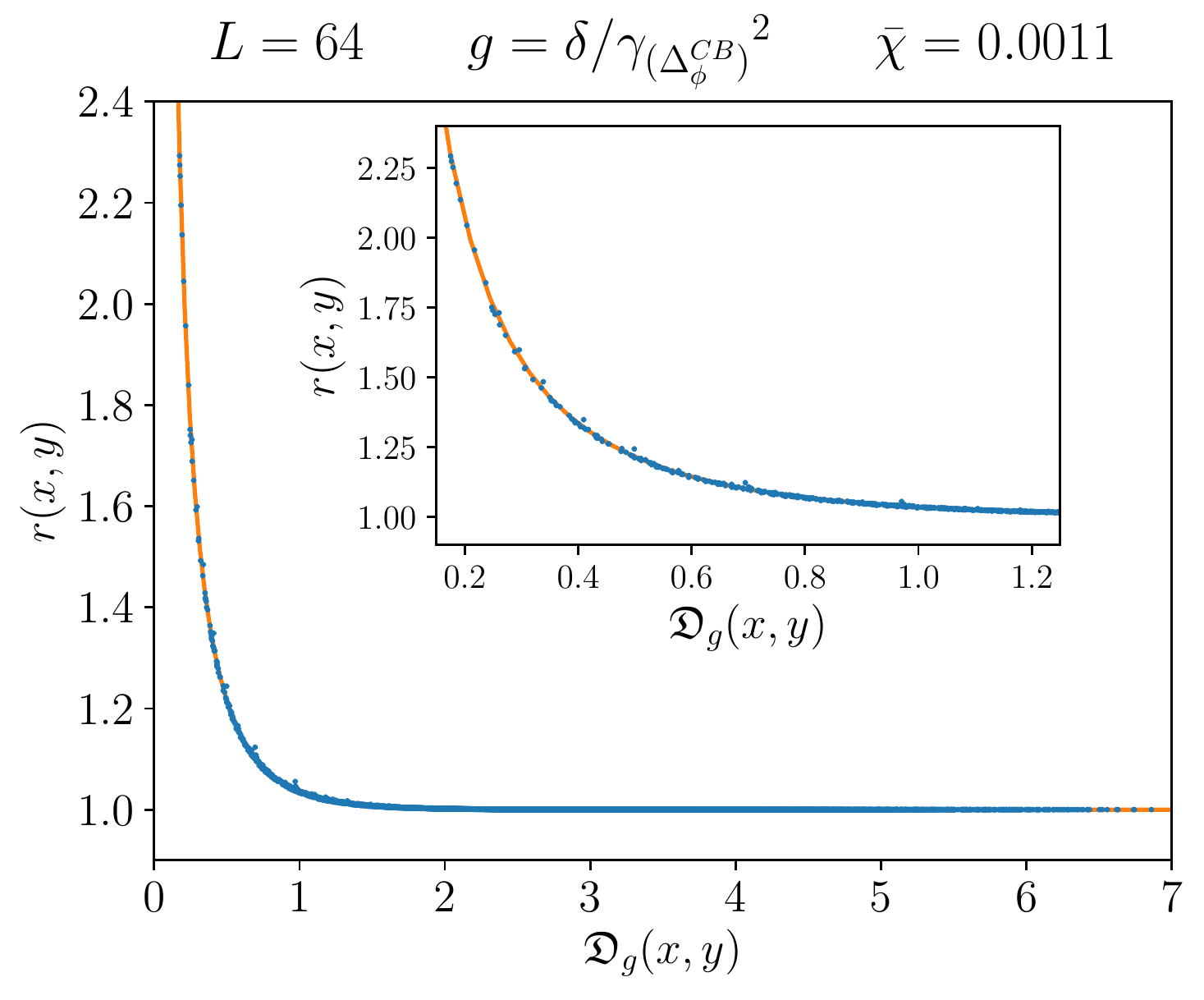}
 \includegraphics[width=.77\textwidth]{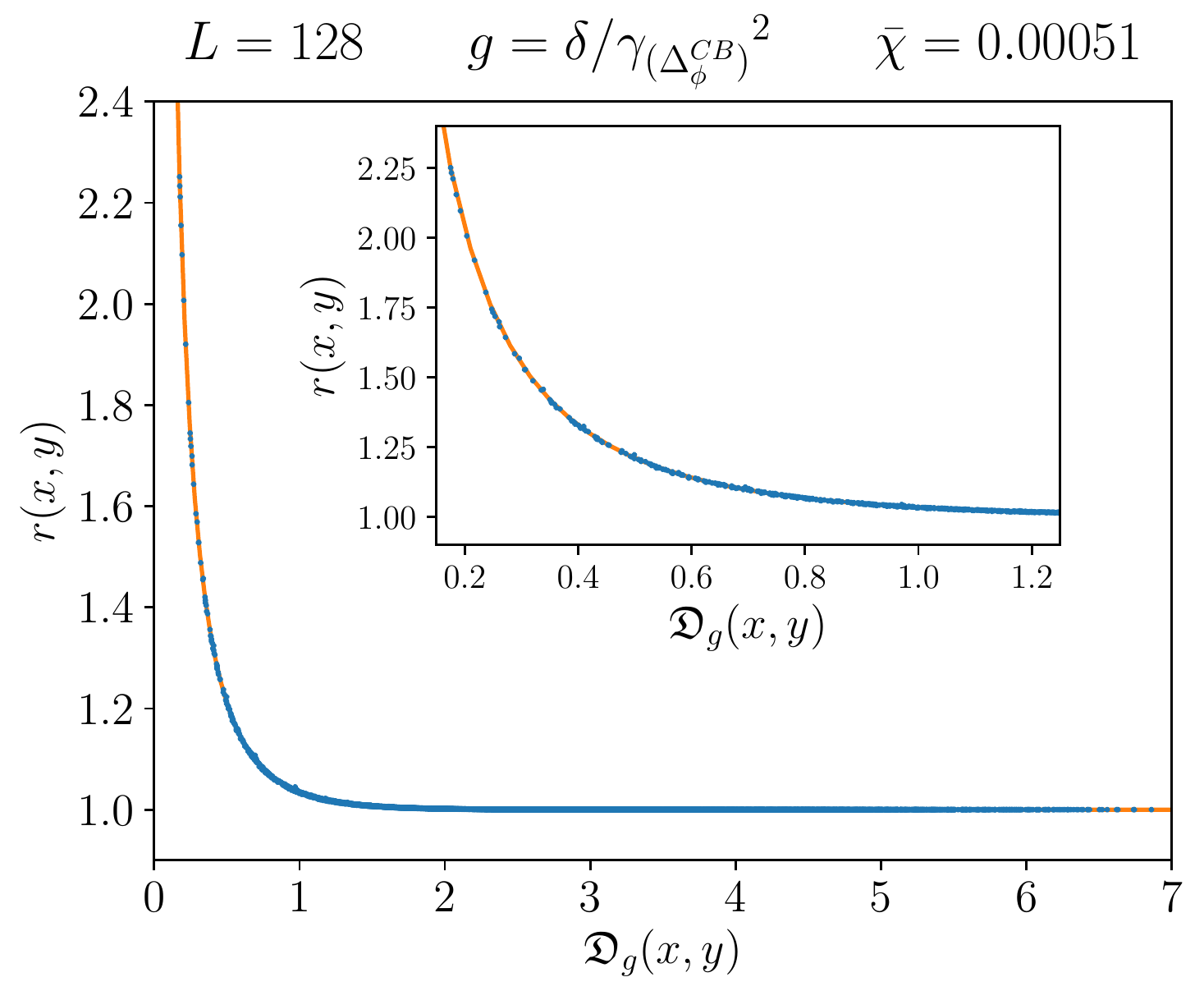}
 \end{center}
 \caption{Correlation ratio $r(x,y)$ defined in Equation~\eqref{ratio}
in terms of distances $\mathfrak{D}$ calculated with
the solution of Fractional Yamabe Equation $\gamma_{(\Delta_\phi^{CB})}$ for the two system sizes $L=64$
and $L=128$. The value of $\bar{\chi}$ is reported
in the upper right corner of the figures. The continuous line is a fitting
function as reported in Appendix~\ref{dataAppendix}.
}\label{fig3}
\end{figure}

As for two-point correlators are concerned we evaluate the ratio:
\begin{equation}\label{ratio}
r(x,y)=\frac{\langle \phi(x) \phi(y)\rangle}{\langle \phi(x)\rangle \langle \phi(y)\rangle},
\end{equation}
and we plot it against the distance 
$\mathfrak{D}_{\delta/{\gamma_{(\Delta_\phi^{CB})}}^2}(x,y)$ calculated with
the metric corresponding to $\Delta_\phi^{CB}$ 
checking whether a collapse of data
points occurs as predicted by
our conjecture~\eqref{conj2ptFYE}. This is
done in Figure~\ref{fig3}.
The collapse is visibly 
good and it gets
better as the system size is 
increased from $L=64$ to $L=128$
with the outliers moving towards
the collapse line.
This has been assessed quantitatively
by calculating a root mean
square of deviations $\bar{\chi}$
from a fitting function
that actually halves as the size
is doubled. In Appendix~\ref{dataAppendix}
details of the analysis of the
two-point correlation functions are reported toghether with
collapse performed with
other metrics.

While the data coming from
two-point correlators fully comply
with our conjecture, the achieved precision
does not allow to rule out 
a geometric description based
on non-fractional Yamabe distance (i.e. the one
based on the solution of Yamabe Equation),
that yields a very similar collapse plot. Again, we consider this 
effectiveness of the description with the Yamabe Equation as a signature 
of the robustness of the devised geometric approach presented in this paper.

\section{Future directions}
The present work aims at giving
a unified description of the
geometrical content of bounded
critical phenomena.
When applied to two-dimensional
systems, it allows to retrieve
from a different perspective 
known results of 
boundary conformal
field theory in $d=2$~\cite{BCFT}.
For higher dimensions 
new predictions for physically relevant observables
were derived. For the 3D Ising model
the comparison between numerical 
results and the predictions
of the theory developed here
is encouraging and it motivates
further efforts along these 
lines. 

We stress that the achieved description
is purely geometric in the sense 
that it depends on the considered
bounded domain $\Omega$
and on the dimension $d$ independently
from the specific model that
one is studying on the domain
-- save for the anomalous dimension $\Delta_\phi$,
which at variance \emph{does} depend on
the model. The scaling exponents 
belong instead to the dynamical,
model dependent,
content of the theory.
One of the advantages of our
approach, and in essence its main feature, 
is to cleanly separate geometry/kinematics 
from dynamics/interaction.
In the well studied case of two dimensions
the former is
simple: since there is no dependence
on $\Delta_\phi$ for $\gamma(x)$,
one geometry fits all models. 
More precisely, 
the metric rendering 
ordinary scalar curvature 
constant and negative makes 
also the fractional Q-curvature
constant washing away
the $\Delta_\phi$ dependence, which is 
not the case for $d>2$.
In $d=2$ the information about 
dynamics is contained in and can be
extracted from the further structure
of the Virasoro algebra.
In this respect many questions are still open.

How different scaling operators other than
the order parameter and
belonging to the same model may have different 
geometries, i.e. different scale factors? 
How do the different
geometries combine to give a consistent
unified framework? Most likely the 
answer to 
these questions 
will lie 
in the close examination of short
distance properties of higher-point
correlation functions in the so-called
operator product expansion which is at the very heart 
of the Conformal Bootstrap approach.
In this light it would be very important
the comparison of the approach presented
here for the energy operator with numerical
simulations in the Ising model.

Moreover it would be highly desiderable
to enlarge the number of models,
shape of bounded domains and boundary
conditions on which our predictions are tested
in 3D and higher.
A natural candidate
would be the XY model that however 
displays a scaling dimension not
differing much from the canonical one. 
Currently work is being
pursued in this direction by 
looking at multicritical points with
different discrete symmetry groups
that appear to have larger deviations
from the canonical dimensions~\cite{aleric}.
Very promising is also the study of percolation models
in $d>2$ with the techniques presented in~\cite{newref1}.
A boundary condition to be
considered is the free boundary condition
pertaining to the so-called ordinary
boundary transition. In addition to considering homogeneous boundary condition
it would be useful to consider also different boundary conditions 
in the same model. These concepts, well developed
in the two-dimensional world (see~\cite{newref2, newref3}  
for recent applications), are central to modern topics such as SLE  
that are still lacking a higher-dimensional counterpart.

We point out that the prediction
for one-point and two-point correlators
in a bounded domain are based on the solution of
a fractional conformally covariant 
differential equation in the same 
domain. To sensibly define the
fractional Laplacian in the
bounded domain with the 
boundary conditions imposed
by criticality, one has to view
the domain as the boundary
of an asymptotically
hyperbolic space living in one more
dimension. This approach
shares some traits with the
AdS/CFT correspondence. It
would be interesting to investigate
the relations between the
geometric approach to criticality 
developed here and the formulation
of statistical mechanics models
at criticality with AdS/CFT
techniques.

This work builds a theory
for critical phenomena deeply rooted in
geometry. Is there an algebraic
counterpart to it? This is to be
intended as a ``critical
Langlands program''. Current work
is being pursued by inspecting
the structure of infinitesimal deformations
of the bounded domain in which Fractional 
Yamabe Equation is studied. 


\section*{Acknowledgements}
The authors wish to thank Mar\'ia del Mar Gonz\'alez and Robin C. Graham for 
providing essential insights into the mathematics needed for
this project. We thank John~L.~Cardy and 
Slava~Rychkov for useful correspondence. 
Discussions with Jacopo Viti and Nicolò Defenu are 
also gratefully acknowledged. 
GG acknowledges important 
discussions with Antonia Ciani.
Moreover GG thanks Universidad Autónoma de Madrid
for hospitality partially funded under BBVA Foundation 
grant for Investigadores y Creadores Culturales 
(actuales becas Leonardo), 2016, during
which crucial stages of the work were performed. 
Early stages of the project were carried on while
GG was visiting ICTP, Trieste.
Both of the authors acknowledge hospitality during 
the ``Disordered systems, random spatial processes 
and some applications'' program held at IHP, Paris, 
during ``Conformal Field Theories and Renormalization 
Group Flows in Dimensions $d>2$'' workshop held at GGI, 
Florence, and a visit to ATOMKI, Debrecen. 
Computational resources were provided by CNR-IOM and SISSA.
GG work is supported by the Deutsche Forschungsgemeinschaft 
(DFG, German Research Foundation) under Germany’s 
Excellence Strategy EXC 2181/1 - 390900948 (the 
Heidelberg STRUCTURES Excellence Cluster).

\appendix

\section{Solution of the Yamabe Equation for the slab domain}\label{YamabeAppendix}
\begin{figure}
 \begin{center}
 \includegraphics[width=.45\textwidth]{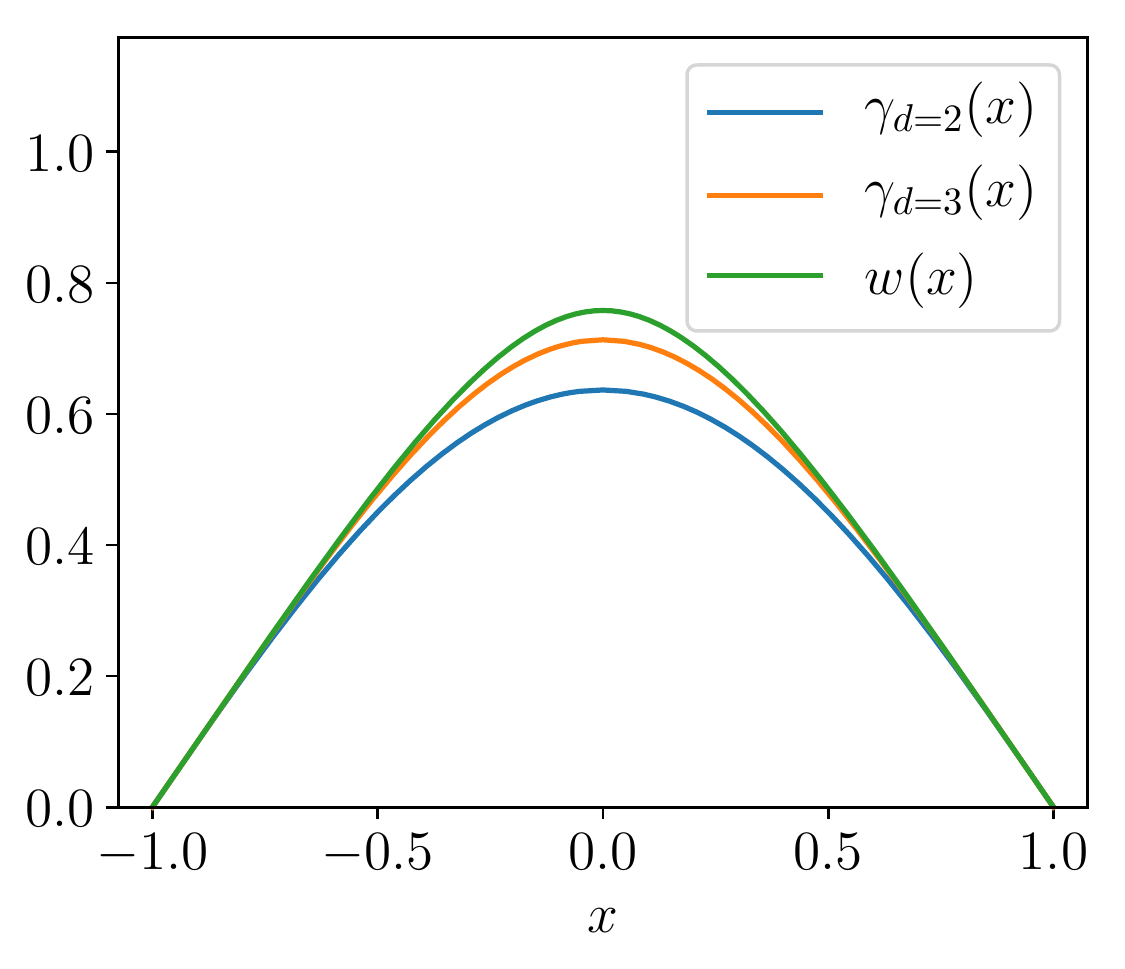}
 \includegraphics[width=.54\textwidth]{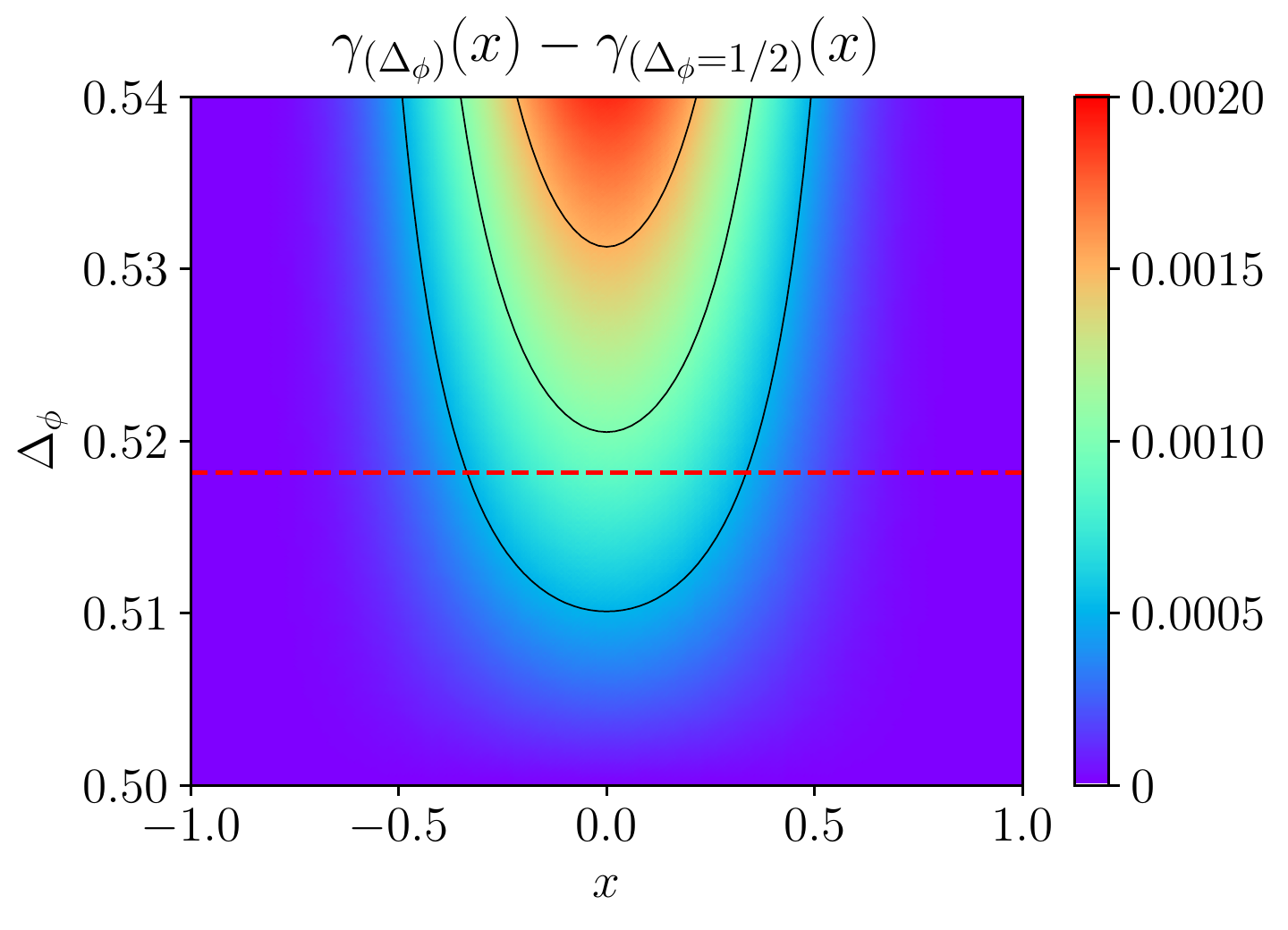}
 \end{center}
 \caption{(Left) Solution of Yamabe problem in the strip ($d=2$) 
 and slab ($d=3$) domain. The additional function $w(x)$ is the
 gauge function used 
 to solve Fractional Yamabe Equation for the slab domain. 
 (Right) Solutions of the Fractional Yamabe Equation 
 as $\Delta_\phi$ is varied in the range $[0.5,0.54]$. The plot actually
 shows deviations from the $\Delta_\phi=1/2$ Yamabe solution (the $d=3$ 
 curve in the left panel). The red dashed line is 
the Conformal Bootstrap value $\Delta_\phi^{CB}$~\cite{boot3}.}\label{IYEandFYE}
\end{figure}

Here we report the solution of Yamabe Equation
for the slab domain in arbitrary dimension $d$ being
defined as $-1<x=x_1<1$ and $x_i \in \mathbb{R}$ for $i=2,\ldots,d$.
The equation for $\gamma(x)$ becomes a nonlinear ordinary differential equation:
\begin{equation}
1 -(\partial_{x} \gamma)^2 + \frac{2}{d}\gamma \partial_{x}^2 \gamma = 0.
\end{equation}
The solution satisfying the appropriate
boundary conditions $\gamma(\pm 1)=0$
can be given in terms of the inverse
function $x(\gamma)$
\begin{equation}
 \pm x(\gamma) = 1 -  \phantom{}_{2}F_{1}\left(\frac{1}{2},\frac{1}{d};1+\frac{1}{d};(\gamma/\gamma_0)^d\right) \gamma,
\end{equation}
where the signs refer to the two symmetric with respect to 
$x\rightarrow -x$ branches and $\gamma_0=\gamma(0)$ is the ($d$-dependent)
metric factor on the symmetry plane $x=0$ of the slab:
\begin{equation}
 \gamma_0=\frac{\Gamma\left(\frac{1}{2}+\frac{1}{d}\right)}{\sqrt{\pi}\,\Gamma\left(1+\frac{1}{d}\right)}.
\end{equation}
$\gamma(x)$ can be explicitly obtained in selected cases: the (trivial) one-dimensional case
$\gamma_{d=1}(x)=\frac{1-x^2}{2}$, the two-dimensional (strip) case 
$\gamma_{d=2}(x)=\frac{2}{\pi}\cos\left(\frac{\pi x}{2}\right)$ while in the
$d\rightarrow \infty$ limit $\gamma_{d\rightarrow \infty}(x)=1-|x|$.
In figure \ref{IYEandFYE} we depict $\gamma_d(x)$ for 
$d=3$, relevant for the
subsequent analysis, and $d=2$ for comparison.

\section{Scattering operators for bounded domains}\label{scatteringAppendix}
A complete discussion of the presented approach, 
including thorough mathematical
justification, will be given in~\cite{MarGori}.

A way to construct the operators with the
required conformal properties is the following.
Consider the domain $\Omega$ equipped with 
the metric $g$ to be the boundary of
a higher dimensional $d+1$ manifold, call it $X$ having
a metric $g_+$.
Denoting the extra coordinate with $y$,
we pose that our original domain $\Omega$ is retrieved
when we set $y=0$ and be regular $\mathrm{d}y|_\Omega\neq0$.
A coordinate $y$ with such properties is called 
a defining function.
Moreover near $\Omega$
the metric $g_+$ should look like $g_+\approx g/y^2$
making our space $(X,g_+)$ asymptotically
 a hyperbolic space. This surface will
be called the conformal infinity.
The metric space $(X,g_+)$ should be an 
Einstein space, that is it has to satisfy
vacuum Einstein field equations 
$\mathrm{Ric}(g_+)+d\,g_+=0$, where $\mathrm{Ric}$
is the Ricci tensor in the $d+1$ dimensional 
space. 

Given a function $f_I$ on $\Omega$ solve the 
following eigenvalue problem for the
function $U$ (defined over $X$):
\begin{equation}
 \begin{cases}
 & (-\Delta_{g_+}) U = \Delta_\phi (d-\Delta_\phi) U \\
 & U = y^{\Delta_\phi} F_I + y^{d-\Delta_\phi} F_O,
 \end{cases}
\end{equation}
where $\Delta_{g_+}$ is the Laplacian for the metric $g_+$,
$F_I$ and $F_O$ are regular functions (the subscripts $I$ and $O$ stand
for input and output respectively)
as $y$ approaches zero and
$F_I|_{y=0}=f_I$. The requirement 
of $(X,g_+)$ to be Einstein guarantees the existence of 
a solution of the above equation (special care has to be
taken when $d-\Delta_\phi$ differs from 
$\Delta_\phi$ by an integer where 
a resonance condition is met and log terms have to be included 
in the expansion). 

The conformal fractional
Laplacian of the function $f_I$
can be read off from the boundary
behavior of $f_O=F_O|_{y=0}$. Indeed we have that
$\mathcal{L}^{(s)}_g f_I = c_s f_O$ 
where $c_s = 2^{2s}\frac{\Gamma (s)}{\Gamma(-s)}$.
The good transformation properties
under conformal changes of this operator
can be seen by choosing a different
defining function, call it $\upsilon$. As the eigenvector
$U$ will be unchanged we have that
\begin{equation}
 U = y^{\Delta_\phi} F_I + y^{d-\Delta_\phi} F_O = \upsilon^{\Delta_\phi} F_I' + \upsilon^{d-\Delta_\phi} F_O' 
\end{equation}
where $F_I'$ and $F_O'$ are respectively input and output data
of another scattering problem. Since around the conformal infinity
the two defining function are linearly related,  
$y=\left.\frac{dy}{d\upsilon}\right|_{\Omega}\upsilon=w(x)^{-1} \upsilon$,
we obtain the desired transformation law:
\begin{equation}
g\rightarrow w(x)^{-2} g, f_I \rightarrow w(x)^{\Delta_\phi} f_I, f_O \rightarrow
 w(x)^{d-\Delta_\phi} f_O.
\end{equation}

\begin{figure}[ht]
\centering
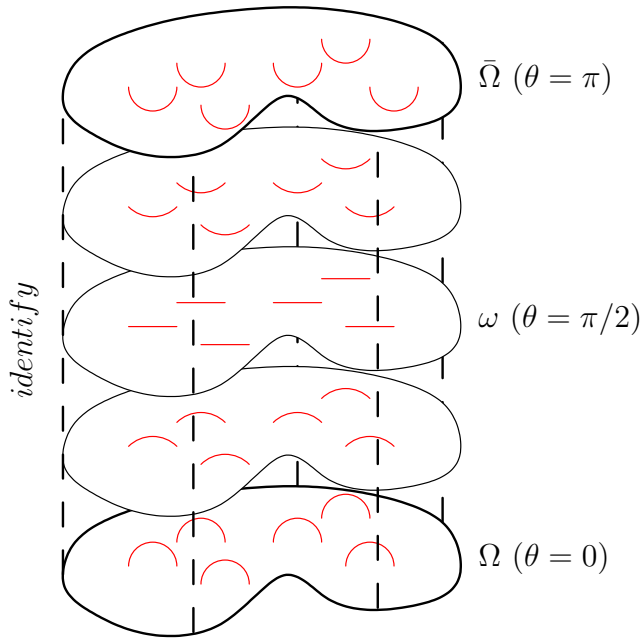
 \caption{Depiction of the extension space $X$. The planes represent
 constant $\theta$ hypersurfaces. The points joined by dotted
 lines should be identified. The red lines are geodesics
 with endpoints on the depicted hypersurfaces.
 The bottom hypersurface is the conformal infinity, the central hypersurface is the 
 totally geodesic surface, and the upper hyperplane is the complement of the 
 conformal infinity.}\label{einstein_manifold}
\end{figure}

The above method first presented in~\cite{graham} has to be adapted 
to the case of bounded domains.
In this case the extension 
metric space $(X,g_+)$ has to fulfill additional 
properties dealing with 
boundaries of $\Omega$, and it 
should be a so-called cornered asymptotic 
hyperbolic space~\cite{mckeown}. There an additional 
surface emerges $\omega$ detached from 
$\Omega$ and sharing the same boundary $\partial \Omega = \partial \omega$.
The surface $\omega$ should
be totally geodesic, that is, geodesics restricted
to $\omega$ should coincide with geodesics in
$X$. Heuristically this can be understood
as a decoupling of what happens inside
$X$ from what happens beyond $\omega$, which acts
as an invisible wall. This construction can 
always be performed, as shown in~\cite{mckeown}, 
provided $\partial \Omega$ is regular 
enough, and the metric $g_+$ can be put in the
canonical form:
\begin{equation}\label{canform}
 g_+ = (\sin \theta)^{-2} (d\theta^2 + g_\theta),
\end{equation}
where the extension variable $\theta$
plays the role of an incidence
angle at the conformal infinity and $g_{\theta=0}=g$. 
In these convenient variables the manifold $X$
is $[0,\pi/2]\times \Omega$ where 
$\theta=0$ is the  surface $\Omega$
and $\theta=\pi/2$ is the totally geodesic
surface $\omega$. The full set of equations 
for the metric becomes:
\begin{equation}\label{cornered}
\begin{cases}
\mathrm{Ric}(g_+)+d\,g_+=0\\
\partial_\theta g_\theta|_{\theta=\pi/2}=0\\
g_\theta^{-1}|_{[0,\pi]\times\partial\Omega}\rightarrow 0
\end{cases}
\end{equation}
supplemented by the regularity of $g_\theta$ for $\theta=0$. 
A cartoon of the space $X$ is represented in Figure~\ref{einstein_manifold}.
As for the extension problem the additional natural
boundary conditions on $U$ are
vanishing normal derivative $\partial_{\mathbf{n}} 
U|_{\partial X\setminus (\Omega
\cup \omega)} = 0$ and also 
$\partial_{\theta} U|_{\partial \omega} = 0$.
With the boundary conditions described 
we have a good candidate for a Neumann
conformally covariant fractional Laplacian\footnote{By extending
farther the space $X$ up to $\theta=\pi$, 
we can arrive at an additional surface $\bar{\Omega}$
which can be regarded as the mirror image of $\Omega$.
The Neumann operator defined and used here amounts to set $U(x,\theta=\pi)=U(x,\theta=0)$.
Other natural boundary conditions are of course conceivable such as 
$U(x,\theta=\pi)=-U(x,\theta=0)$ (Dirichlet-like) and more 
generally $U(x,\theta=\pi)=\alpha U(x,\theta=0)$ (Robin type).}.

We now provide an example
where the fractional Laplacian
can be calculated and where 
we verify that the hyperbolic 
metric indeed solves
the non-compact fractional Yamabe problem. 
Consider the upper half hyperspace $\mathbb{R}_+^d$ in 
$d$ dimensions and view it 
as the boundary of the space $X=[0,\pi/2]\times\mathbb{R}_+^d$;
the extension coordinate being $\theta$.
The metric
\begin{equation}
 g_+ = (\sin \theta)^{-2} (d\theta^2 + g_\theta)= (\sin \theta)^{-2} (d\theta^2 + dx^2/x_d^2)
\end{equation}
satisfies the (cornered) Einstein vacuum equation~\eqref{cornered}.
If we plug in the function 
\begin{equation}
U=\tau^{\Delta_\phi} \eta(\Delta_\phi,\tau)+\frac{\tau^{d-\Delta_\phi}}{c_s}\frac{\Upsilon(\Delta_\phi)}{\Upsilon(d-\Delta_\phi)} \eta(d-\Delta_\phi,\tau), 
\end{equation}
where $\tau=\tan(\theta)$ and 
$\eta(\Delta_\phi,\tau)={}_2F_1(\frac{\Delta_\phi}{2},\frac{1+\Delta_\phi}{2};1-\frac{d}{2}+\Delta_\phi;-\tau^2)$, we see that the scattering problem is 
actually satisfied with the correct boundary conditions.
What we have just described proves that 
$$\mathcal{L}^{(\frac{d}{2}-\Delta_\phi)}_{\delta/x_d^2} (\mathbbm{1}) = 
\frac{\Upsilon(\Delta_\phi)}{\Upsilon(d-\Delta_\phi)} (\mathbbm{1}),$$
meaning that the hyperbolic metric in $\mathbb{R}_+^d$
has constant fractional Q-curvature $R_{\delta/x_d^2}^{(\frac{d}{2}-\Delta_\phi)} = -1$ (cfr. Equation~\eqref{fracQR}).
The performed computation for $d=2$ can be adapted to
any regular enough domain via a conformal mapping 
viewed as an isometry between models of the hyperbolic
plane. For $d>2$ it allows to retrieve
the metric factor just for the hyperball domain
which is the only one isometric to the hyperbolic
space.
For other domains we will have to proceed
as specified in the next Appendix.

%
\section{Spectral solution of the Fractional Yamabe Equation in the slab geometry}\label{spectralFYEAppendix}
\begin{figure}
 \begin{center}
 \includegraphics[width=\textwidth]{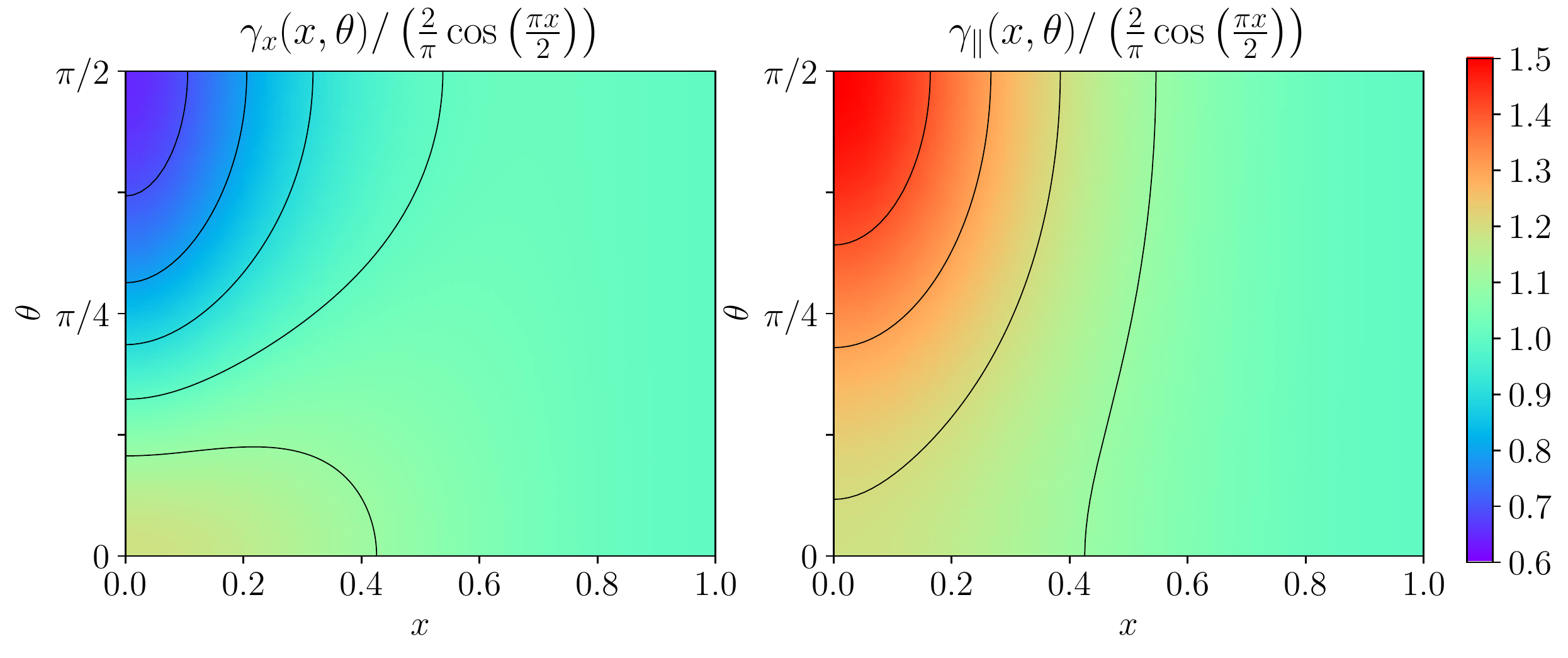}
 \end{center}
 \caption{Solution of Einstein equations
 as specified by the functions $\gamma_x(x,\theta)$
 and $\gamma_\parallel(x,\theta)$
 in the extension space above the slab. The 
 domain shown is $x\in[0,1]$ and $\theta\in[0,\pi/2]$.
 The regions $x\in[-1,0]$ and $\theta\in[\pi/2,\pi]$
 can be obtained by reflection. To improve readability, 
 the functions have been divided by $\frac{2}{\pi}\cos
 \left(\frac{\pi x}{2}\right)$.
 }\label{EEcontour}
\end{figure}

Further details of this solution will also
be presented in~\cite{MarGori}.
Before defining the conformal
fractional Laplacian for the slab domain
a solution to Einstein Equations has to
be found. For the slab case
the metric can be put in the 
canonical form~\eqref{canform}
with a diagonal metric:
\begin{equation}
 g_+ = (\sin \theta)^{-2} \left[d\theta^2 + dx^2/\gamma_x(x,\theta)^2 + (dx_2^2+dx_3^2)/\gamma_\parallel(x,\theta)^2\right]
\end{equation}
(remember that $x=x_1\in[-1,1]$ denotes
the transverse direction of the slab).
With this Anstatz a solution of~\eqref{cornered}
has been found.
In Figure~\ref{EEcontour} the functions
$\gamma_x(x,\theta)$ and $\gamma_\parallel(x,\theta)$
are shown. On the conformal
infinity $\theta=0$ they do by construction coincide, 
thus providing a good metric 
to calculate conformally invariant Laplacians
in the flat conformal class.
The function $\gamma_x(x,0)=\gamma_\parallel(x,0)=w(x)$
shown in Figure~\ref{IYEandFYE} will 
specify the actual gauge in which 
computations will be performed.

Because of the outlined structure of the solution
of the extension problem
near the conformal infinity, the following form will
be assumed for $U$:
\begin{align}
 U &= \sin(\theta)^{\Delta_\phi} u = \nonumber\\ 
 &= \sin(\theta)^{\Delta_\phi} \left( \sum_{i=0,\, i\, \mathrm{even}}^{N_\theta}
 \mathcal{F}_i(x) \sin(\theta)^{2i}
 + \sum_{i=1,\, i\, \mathrm{odd}}^{N_\theta}\mathcal{F}_i(x) \sin(\theta)^{2s+i-1} \right).
\end{align}
The function $\mathcal{F}_i(x)$  
are even functions that will be represented
as linear combinations of $N_x$ even Chebyshev polynomials $T_{2 j}(x)$:
\begin{equation}
 \mathcal{F}_i(x) = \sum_{j=0}^{N_x} T_{2 j}(x) \psi_{i,j}.
\end{equation}
The extension partial differential equation 
problem will be evaluated on a set 
of $N_\theta \times N_x$ collocation
Gauss-Lobatto points given by 
$\theta_i=\frac{\pi}{2}\left[1-\cos\left(\frac{\pi i}{2 N_\theta}\right)\right]$,
$x_j=\cos\left(\frac{\pi j}{2 N_x}\right)$ with $i=1,\ldots,N_\theta$
and $j=1,\ldots,N_x$, reducing it to a solution of a matrix equation.
The boundary conditions are given by $\partial_x u = 0$ on $x=\pm 1$
and $\partial_\theta u = 0$ on $\theta=\pi/2$.
The input data (the function of which 
we are calculating the fractional Laplacian) 
is given by the $\mathcal{F}_0(x)$, while the
output (the fractional Laplacian) 
is given by $\mathcal{L}^{(s)}_{\delta/w^2}[\mathcal{F}_0](x)=c_s \mathcal{F}_1(x)$.

Let us turn to the nonlinear eigenvalue 
problem contained in~\eqref{FYEcovariant}.
By numerical experimentation, the framework
proves more stable if we work 
with the inverse of the fractional Laplacian 
$\mathcal{I} = \frac{\Upsilon(\Delta_\phi)}{\Upsilon(d-\Delta_\phi)}
[\mathcal{L}^{(s)}_{\delta/w^2}]^{-1}$.
The form we have actually considered is:
\begin{equation}
\mathcal{E}[\rho] = \mathcal{I}[\rho]-\rho^{\frac{\Delta_\phi}{d-\Delta_\phi}} = 0,
\end{equation}
where $\rho=\left(\gamma/w\right)^{-d+\Delta_\phi}$.
The above equation has been
solved by looking for a minimum 
of $\sum_{j=0}^{N_x}|\mathcal{E}[\rho](x_j)|^2$.
This minimisation yields, in a stable way, a
very small value (on the order of $10^{-24}$)
signaling that the equation is 
satisfied to a very high accuracy.
The chosen collocation grid for the numerical solution 
is $N_x=N_\theta=20$. Since the values of
$\Delta_\phi$ of interest to us 
are around $\Delta_\phi\approx0.52$, we 
solved the Fractional Yamabe Equation on a Gauss-Lobatto grid of $12$ values
in $(0.5,0.54]$ allowing us to obtain
reliable solutions for the fractional Yamabe problem in the slab in this 
range via Chebyshev interpolation.
The solutions turn out to differ only
slightly from the solution of Yamabe Equation and connect smoothly to it when
$\Delta_\phi\rightarrow 1/2$ (that is $s\rightarrow1$).
The deviations from the Yamabe Equation solution 
for the slab domain are shown in
the right panel of Figure~\ref{IYEandFYE}.

\section{Monte Carlo experiments}\label{MCAppendix}
The model we simulate has the Hamiltonian:
\begin{equation}
 \mathcal{H}=-\beta\sum_{<i,j>} s_{i} s_{j} + D \sum_{i} 
 s_i^2,
\end{equation}
where the spins $s_{i}$ are located
on a cubic lattice $i=(i_1,i_2,i_3)$ and assume the
three values $s_i=-1,0,1$.
$<i,j>$ denote nearest 
neighbors. The parameters 
take the values $\beta=0.387721735$ and $D=0.655$ 
for the system to be at the critical point in their current
best estimates~\cite{Hasenbusch0}.
We expect that, going to larger systems,
more precise determination of the
critical point will be in order.
The geometry we consider is a three-dimensional 
slab, with $0\leq i_1 \leq L$ and $0\leq i_2, i_3\leq L_\parallel$. 
$L$ should be made larger and larger
(lattice spacing is set to one).
Of course we will be simulating a finite system
approximating the specified geometry.
The transverse direction will have $L+1$ sites 
(out of which $L-1$ will be left free to vary)
and the parallel ones $L_\parallel$ sites with 
periodic boundary conditions along the parallel directions.
The boundary sites in the transverse
direction will be fixed to one.
This boundary condition, known as extraordinary, 
is imposed in order to develop a nonzero
order parameter. In order to make the
finiteness of the parallel directions
less relevant, but the system size still
tractable, we will choose $L_\parallel= 6 L$
which we checked to be large enough to  
cancel the dependence on $L_\parallel$ in the measured observables.
The sizes considered are $L = 32, 48, 64, 96, 128, 192$
reaching a maximum of $191\cdot 1152^2 \approx 2.5\cdot 10^8$
free sites. We expect that in order to
perform simulations for larger sizes would
require a more precise determination
of the critical parameters with respect
to the ones in~\cite{Hasenbusch0}.
The algorithm used is the one 
described in~\cite{Hasenbusch1}. Moreover,  
in order to reduce statistical fluctuations an
analytical variance reduction technique has been used
similarly to what done in~\cite{Hasenbusch2}
in a numerical way:
it amounts to summing over states of the spin 
under consideration and its nearest neighbors 
exactly.

The number of samples collected after a 
suitably long thermalisation stage of $10^4$ 
MC steps is on
the order of $10^6$ samples. Of course
autocorrelation reduces the number 
of independent samples. This has been dealt
with by blocking techniques~\cite{newref4}.

\section{Data analysis}\label{dataAppendix}

The variance reduced one-point data
have been averaged along parallel
directions to reduce statistical
fluctuations.
The fitting procedure with function 
\eqref{onepointfit} has been adapted
to the raw data in the following 
way. The points in 
the center of the slab are 
more sensitive to the functional 
form of the universal scaling function
which however has a rather generic
parabolic shape.
The points near the boundary
on the other side, while
allowing for direct access to the 
critical exponents, are more affected
by finite size effects.
In order to get the most out
of our data a weighting window function
has been applied to the data: 
it is one in the slab center,
decays linearly with a width of two sites
and it is zero beyond that point.
The position of this window has been
adapted in a continuous way such that our theory
is consistent with a $p$-value
of $95\%$. These are the data reported 
in Table \ref{table2}.
The required windowing becomes 
smaller and smaller as the 
size is increased with the 
center of the linear part of the
window being located at $x\simeq 0.9$ for
the largest size $L=192$.
We remark that if a similar analysis
is carried out with the, not theoretically justified
but anyway seemingly sensible,
magnetization profile for the 
strip $\gamma_{d=2}(x)^{-\Delta_\phi}$ 
the data yield the incorrect estimate
for $\Delta_\phi\simeq0.8$
with a heavy windowing of
the data keeping only the central 
half of the points.
This clearly rules out the description
with $\gamma_{d=2}$.

As for two-point data, we
recorded the spin-spin correlation functions 
for all the possible
distinct pairs $(i,j)$
of points in the following
set: $i_1=m_1(L/16)$, $i_2=0$, $i_3=0$ 
and $i_1=n_1(L/16)$, $i_2=n_2(L/16)$, $i_3=n_3(L/16)$
with $m_1, n_1=1, \ldots, 15$
and $n_2, n_3=0, \ldots, 14$.
This yields a total of $7672$ 
independent correlators
since coinciding points
have been excluded and pairs symmetric under 
reflection have been put together.
Average over parallel
directions has obviously been
performed.
For these measurements 
the data set is approximately
$10$ times smaller than the
one collected for one-point
functions.

\begin{figure}
 \begin{center}
 \includegraphics[width=.77\textwidth]{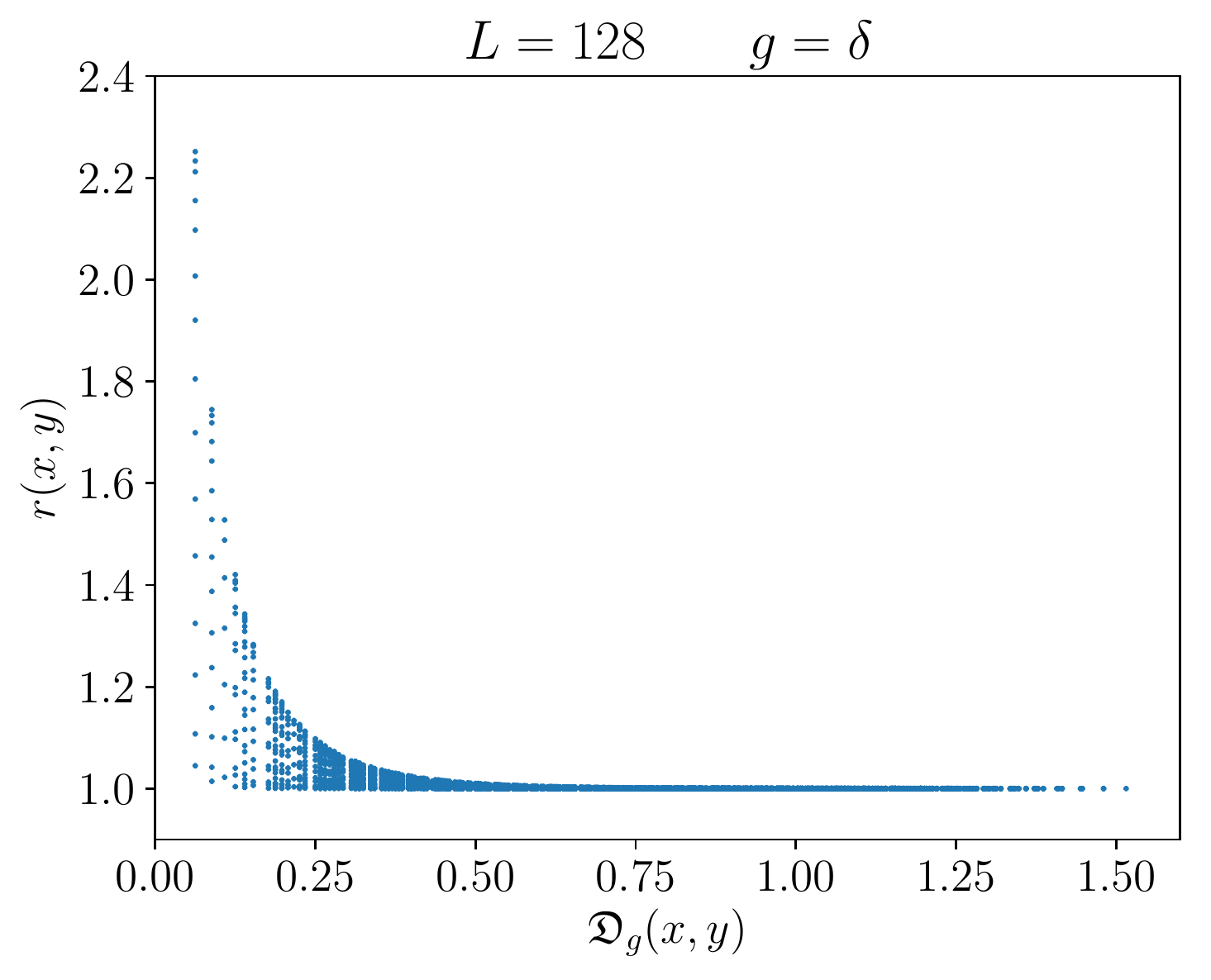}
 \end{center}
 \caption{Correlation ratio $r(x,y)$ defined in Eq.~\eqref{ratio} 
in terms of the euclidean distance $\mathfrak{D}_{\delta}$ for $L=128$.
}\label{wrongcollapse}
\end{figure}

\begin{figure}
 \begin{center}
 \includegraphics[width=.77\textwidth]{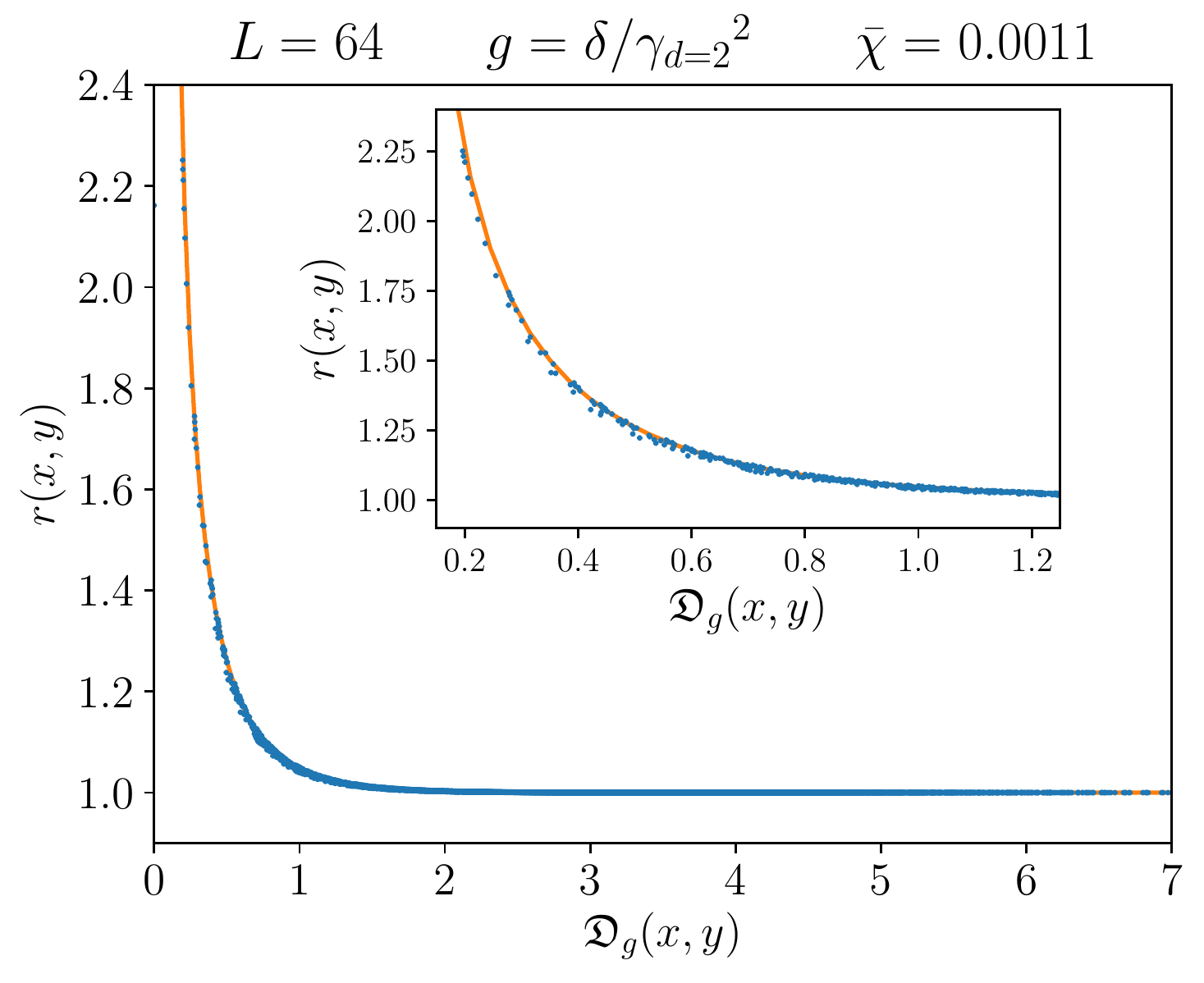}
 \includegraphics[width=.77\textwidth]{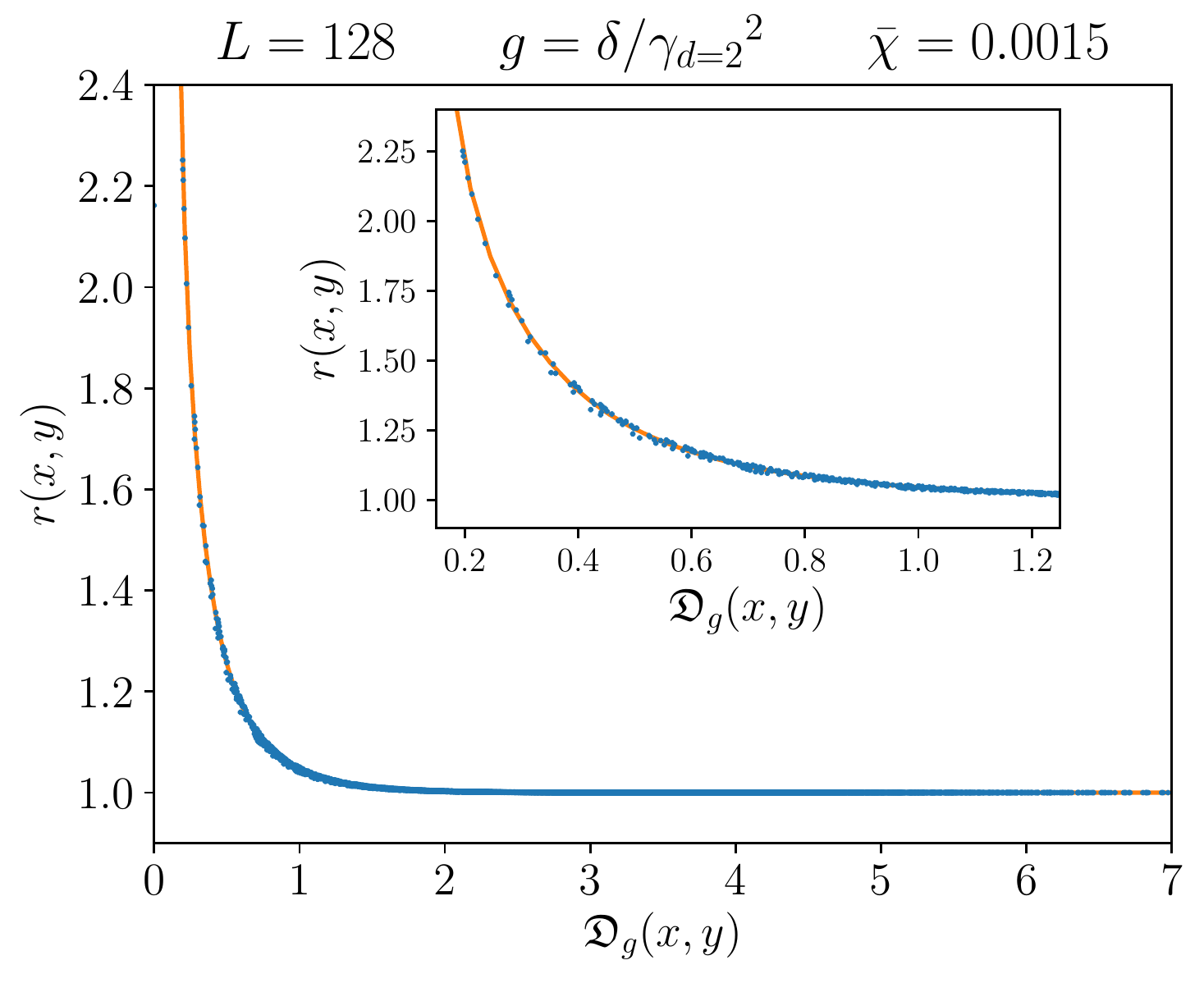}
 \end{center}
 \caption{Correlation ratio $r(x,y)$ 
in terms of distances $\mathfrak{D}$ calculated
with the strip distance $\gamma_{d=2}=\left(\frac{2}{\pi}\right) 
\cos\left(\frac{\pi x}{2}\right)$ for the two system sizes $L=64$
and $L=128$. The value of $\bar{\chi}$ is also reported.
The continuous line is the fitting function $f$, see text for details.}
\label{fig4}
\end{figure}

For each of these couple of points
the distance has been calculated
and plotted against ratio~\eqref{ratio}.
In order to see how a collapse
can go wrong let us check
a case where it should not work
from the outset. While rotational
symmetry is broken we may nonetheless
plot as a check our ratio against euclidean
distance: this yields, as expected, the very poor collapse 
shown in Figure \ref{wrongcollapse}.
For the (more difficult to handle) metric $\delta/\gamma_{(\Delta_\phi)}^2$
the calculation of the distances has been
performed numerically taking advantage of the 
geodesics functionality present 
in the program Surface Evolver~\cite{brakke}.
In order to assess the goodness
of the collapse, we fit $r$
as a function of $\mathfrak{D}_g$
with the function $f(x)=1+\sum_{i=1}^3 a_i e^{-b_i x}$. 
This yields a reasonable description of the data
that will be taken with no errors since they
are much smaller than the observed spread.
The figure of merit will be 
the root mean square of deviations
from the fitting function:
\begin{equation}
 \bar{\chi} = \sqrt{\frac{\left[r-f(\mathfrak{D}_g)\right]^2}
{n_{\mathrm{d.o.f.}}}}, 
\end{equation}
where $n_{\mathrm{d.o.f.}}=7666$ is the number of degrees of freedom.
Let us see quantitatively whether a collapse
occurs with the, already ruled out 
by one-point correlator analysis, strip metric factor
$\gamma_{d=2}$. Results are shown
in Figure~\ref{fig4}. While yielding 
a reasonable collapse, we remark that
points with $\mathfrak{D}\simeq 1$
have a considerable spread and, more
importantly the value of $\bar{\chi}$
\emph{increases} with the system size,
ruling out again $\gamma_{d=2}$ as the correct
metric factor.

\end{document}